\documentclass{article}
\pdfoutput=1
\usepackage{jheppub}
\usepackage{pstricks}
\usepackage[latin2]{inputenc}
\usepackage{epic}
\usepackage{latexsym}
\usepackage{epsf}
\usepackage{epsfig}
\usepackage{amssymb,amsmath}
\usepackage{amsfonts}
\usepackage{dsfont}
\usepackage{graphicx}
\usepackage{epstopdf}
\usepackage{graphics}
\usepackage{subfigure}
\usepackage{multirow}
\usepackage{amssymb} 
\usepackage{ulem}
\usepackage{amsmath}
\usepackage{bbm} 
\usepackage{mathrsfs}   
\usepackage{xspace}

\newcommand{\beq}{\begin{equation}} 
\newcommand{\eeq}{\end{equation}} 
\newcommand{\ba}{\begin{array}}  
\newcommand{\ea}{\end{array}} 
\newcommand{\bea}{\begin{eqnarray}}  
\newcommand{\eea}{\end{eqnarray} }  
\newcommand{\be}{\begin{eqnarray}}  
\newcommand{\ee}{\end{eqnarray} }  
\newcommand{\bal}{\begin{align}}
\newcommand{\eal}{\end{align}}   
\newcommand{\bi}{\begin{itemize}}  
\newcommand{\ei}{\end{itemize}}  
\newcommand{\ben}{\begin{enumerate}}  
\newcommand{\een}{\end{enumerate}}  
\newcommand{\bc}{\begin{center}}
\newcommand{\ec}{\end{center}} 
\newcommand{\bt}{\begin{table}}
\newcommand{\et}{\end{table}}  
\newcommand{\btb}{\begin{tabular}}
\newcommand{\etb}{\end{tabular}}

\newcommand{\eVdist}{\kern-0.06em}

\begin{document}


 \vspace*{-3.0cm}
 \begin{flushright}
 OUTP-17-01P\\
 DESY-17-008\\
 KA-TP-01-2017 
\end{flushright}

\title{Revisiting fine-tuning in the MSSM }

\author[a]{Graham G.~Ross}
\affiliation[a]{Rudolf Peierls Centre for Theoretical Physics, University of Oxford,1 Keble Road, Oxford OX1 3NP, UK}
\emailAdd{g.ross1@physics.ox.ac.uk} 

\author[b]{Kai Schmidt-Hoberg}
\affiliation[b]{DESY, Notkestrasse 85, D-22607, Hamburg, Germany}
\emailAdd{kai.schmidt-hoberg@desy.de} 

\author[c,d]{Florian Staub}
\affiliation[c]{Institute for Theoretical Physics (ITP), Karlsruhe Institute of Technology, Engesserstra{\ss}e 7, D-76128 Karlsruhe, Germany}
\affiliation[d]{Institute for Nuclear Physics (IKP), Karlsruhe Institute of Technology, Hermann-von-Helmholtz-Platz 1, D-76344 Eggenstein-Leopoldshafen, Germany}
\emailAdd{florian.staub@kit.edu}

\abstract{
We evaluate the amount of fine-tuning in constrained versions of the minimal supersymmetric standard model (MSSM), with different boundary conditions at the GUT scale.
Specifically we study the fully constrained version as well as the cases of non-universal Higgs and gaugino masses. We allow for the presence of additional non-holomorphic soft-terms which we show further relax the fine-tuning. Of particular importance is the possibility of a Higgsino mass term and we discuss possible origins for such a term in UV complete models.
We point out that loop corrections typically lead to a reduction in the fine-tuning by a factor of about two compared to the estimate at tree-level, 
which has been overlooked in many recent works.  Taking these loop corrections into account, we discuss the impact of current limits from SUSY searches and dark matter on the fine-tuning. Contrary to common lore, we find that the MSSM fine-tuning can be as small as 10 while remaining consistent with all experimental constraints. If, in addition, the dark matter abundance is fully explained  by the neutralino LSP, the fine-tuning can still be as low as $\sim 20$ in the presence of additional non-holomorphic soft-terms.
We also discuss future prospects of these models and find that the MSSM will remain natural even in the case of a non-discovery in the foreseeable future.}

\maketitle

\section{Introduction}
\label{s.Intro}
The discovery of the Higgs with a mass of about 125~GeV \cite{Aad:2012tfa,Chatrchyan:2012xdj} is, to date, the biggest success of the large hadron collider (LHC). In contrast, there has not been any evidence for new physics. This puts very strong constraints on the masses of new coloured particles as predicted, for instance, by supersymmetry (SUSY); working with very simplified assumptions, it is possible to exclude gluinos and first/second generation squarks nearly up to 2~TeV \cite{ATLAS-CONF-2016-052,ATLAS-CONF-2016-054,ATLAS-CONF-2016-078,Xie:2223502}. These experimental results raise the question of how natural are the simplest versions of supersymmetry such as the minimal supersymmetric standard model (MSSM). It has become common lore that a strong tension between the LHC results and `natural SUSY' is now the case; for example in the Constrained MSSM  it was found, because of the rather heavy Higgs mass, that one needs a fine-tuning (FT) of 300 or more to accommodate correct electroweak symmetry breaking (EWSB)\cite{Cassel:2011tg}. This has led to increasing interest in less constrained versions of the MSSM \cite{Ghilencea:2012gz,vanBeekveld:2016hug} and also in other SUSY models such as those with singlet extensions which can increase the Higgs mass. It was found that these models can still be considered as mildly fine-tuned  even in fully constrained scenarios, see e.g. Refs.~\cite{Ross:2011xv,Ross:2012nr,Kaminska:2014wia} and references therein. 

However, even in the MSSM, the situation is not as clear as often claimed; the main source of FT is usually the $\mu$-term in the superpotential of the MSSM which gives mass to the Higgsinos and simultaneously contributes to the Higgs masses. As a result natural SUSY is usually associated with light Higgsinos, not heavier than a few hundred GeV \cite{Baer:2012up,Baer:2012uy,Baer:2012cf,CahillRowley:2012rv,Feng:2012jfa,Kang:2012sy,Baer:2013gva,Baer:2013vpa,Kowalska:2013ica,Baer:2014ica}. It has been recently pointed out that this is not strictly correct because of the possibility of an additional source of  Higgsino mass \cite{Ross:2016pml}; if the Higgsinos gain mass via a soft-breaking term $\mu'$, which does not affect the FT very much, it is possible to get 1~TeV Higgsinos without a large FT penalty. 
We show that such a large mass can arise through large radiative corrections involving  Higgs portal couplings to SUSY breaking in a hidden sector. It is also possible in sequestered SUSY breaking that radiative effects generate a natural cancellation between the $\mu$ term and the soft Higgs mass terms that leaves only an Higgsino mass \cite{Perez:2008ng}.  Moreover, if one relaxes the requirement of the unification of the gaugino mass terms at the GUT scale, regions of parameter space exist with only mild fine-tuning \cite{Antusch:2012gv,Kaminska:2013mya}. This is a natural possibility in many GUT models where the gaugino mass terms get split by the breaking of the enhanced gauge group and also in string theory models (see \cite{Horton:2009ed} and \cite{Kaminska:2013mya} and references therein). 

The aim of this work is to perform a study of the fine tuning in constrained versions of the MSSM extended by non-holomorphic Higgsino mass and other soft-terms and to discuss the impact of gaugino mass hierarchies at the GUT scale. In the course of this work we found that the loop-corrections to the FT can be very important and typically reduce the FT prediction by a factor of 2.\footnote{Recently a large impact of higher-order corrections has been found also for other fine-tuning measures~\cite{Buckley:2016tbs}.} This disagrees with the treatment of loop corrections in Ref.~\cite{Baer:2012cf} and with a series of recent papers that follow this treatment. We discuss the origin of this disagreement and give the full prescription of how to deal with loops when calculating FT. To take account of these corrections we have added a section revising the results for the MSSM.

This paper is organised as follows: in Section~\ref{sec:loopFT} we discuss the loop corrected FT calculation and compute the improvement compared to the previous calculations. In Section~\ref{sec:NHSSM} we give a short introduction to the MSSM extended by non-holomorphic soft-terms and we analyse possible origins of a large Higgsino mass term, both of holomorphic and non-holomorphic origin. In Section~\ref{sec:results} we present the results of  exhaustive numerical scans for the FT in the MSSM, constrained by the latest experimental results.  We conclude in Section~\ref{sec:conclusions}.

\section{Revisiting the fine-tuning calculation in the MSSM including loop corrections}
\subsection{Fine-tuning measure at tree-level}
\label{sec:loopFT}
Let us start our discussion of the MSSM fine tuning at tree-level, in order to gain some intuition and to connect to most results in the literature.
After electroweak symmetry breaking (EWSB) the Higgs potential in the MSSM is given by
\begin{equation}
V =  \frac12 m_{h_d}^2 v_d^2 + \frac12  m_{h_u}^2 v_u^2 + \frac12 \mu^2 (v_d^2+v_u^2) - \frac12 (B_\mu v_d v_u + h.c) + \frac{1}{32} (g_1^2+g_2^2)(v_d^2-v_u^2)^2
\end{equation}
from which one derives the minimum conditions (tadpole equations) 
\begin{equation}
T_i=\frac{\partial V}{\partial v_i} \equiv 0  \hspace{1cm} i=d,u
\end{equation}
as
\begin{align} 
\label{eq:tad1}
T_d &= - v_u \Re\Big(B_{\mu}\Big) + \frac{1}{8} \Big(g_{1}^{2} + g_{2}^{2}\Big)v_d \Big(v^2_d - v^2_u \Big)+ v_d \Big(m_{h_d}^2 + |\mu|^2\Big) \;,\\ 
\label{eq:tad2}
T_u &= \frac{1}{8} \Big(g_{1}^{2} + g_{2}^{2}\Big)v_u \Big(v_{u}^{2} - v_{d}^{2}  \Big) - v_d {\Re\Big(B_{\mu}\Big)}  + v_u \Big(m_{h_u}^2 + |\mu|^2\Big) \;.
\end{align} 
All SUSY parameters are understood as running $\overline{\text{DR}}$ quantities at the SUSY scale. 
As usual, we define $\tan\beta=\frac{v_u}{v_d}$ and note that $B_\mu$ is proportional to the mass $M_A$  of the pseudo-scalar Higgs boson. 
In the decoupling limit ($M_A \to \infty$) and for $\tan\beta\to \infty$ one finds the simple relation
\begin{align}
v \mu^2 + v m_{h_u}^2 + \frac18 (g_1^2 + g_2^2) v^3 = 0 
\end{align}
which is often presented in the form
\begin{equation}
\label{eq:tree}
\frac12 M_Z^2 = - \mu^2 - m_{h_u}^2  \;.
\end{equation}
This makes the origin of the (little) hierarchy problem within the MSSM apparent: the r.h.s. contains terms which are naturally $\mathcal{O}(M_{\rm SUSY})$, the SUSY breaking scale. Thus, in order to obtain the measured value of $M_Z$ there must be a cancellation between these terms which demands a certain level of tuning. 
There are different measures to quantify the amount of fine-tuning $\Delta_{FT}$. A widely used one is the sensitivity measure proposed in Refs.~\cite{Ellis:1986yg,Barbieri:1987fn}
\begin{align} 
\label{eq:measure}
\Delta \equiv \max {\text{Abs}}\big[\Delta_{p}\big],\qquad \Delta _{p}\equiv \frac{\partial \ln
  v^{2}}{\partial \ln p} = \frac{p}{v^2}\frac{\partial v^2}{\partial p} \;.
\end{align}
Here, $p$ are the independent parameters of the model, and the quantity $\Delta^{-1}$ gives a measure of the accuracy to which independent parameters must be tuned to get the correct electroweak breaking scale\footnote{The fine tuning measure can be related to a factor in a likelihood fit to the data. Interpreting this in  a probabilistic sense fine tuning of $\mathcal{O}(50)$ adds about 0.5 to the value of $\chi^2/$degree of freedom in a SUSY fit to data \cite{Ghilencea:2012qk}.}. Applying this measure to eq.~(\ref{eq:tree}), one finds 
\begin{equation}
 \frac{\partial \ln
  v^{2}}{\partial \ln p_i} =\frac{\partial\ln M_Z^2}{\ln p_i} = 2 \frac{p_i^2}{M_Z^2} \left(-\frac{\partial \mu^2}{\partial p_i^2}-\frac{\partial m_{h_u}^2}{\partial p_i^2}\right)
\end{equation}
Using $p^2=\{\mu^2,m_{H^2_u}\}$ the very naive estimate for the fine-tuning is found to be
\begin{equation}
\Delta_\mu = -\frac{2 \mu^2}{M_Z^2} \,, \hskip1cm \Delta_{m_{h_u}^2} = -\frac{2 m^2_{h_u}}{M_Z^2} \;.
\label{tree}
\end{equation}
Thus, a small FT needs moderately small $|\mu|$ and $|m_{h_u}^2|$ at the low scale.

In the following, we do not discuss the FT as function of the parameters at the SUSY scale, but consider a high-scale model in which the fundamental parameters are defined at the scale of grand unification (GUT scale).  The reason is that in UV complete models it may be incorrect to treat all parameters as independent at the SUSY scale because  correlations among parameters are usually present. These correlations can significantly affect the calculated fine-tuning. For instance if SUSY breaking leads to degenerate soft scalar masses there is a cancellation between the tree level and radiative contributions to the Higgs mass that leads to a reduction of the sensitivity of the Higgs mass to the initial scalar masses, the so-called `focus point' \cite{Chan:1997bi,Feng:1999mn,Feng:1999zg,Feng:2000gh}. As a result the fine-tuning measure is significantly reduced. It has been pointed out in Refs.~\cite{Choi:2005hd,Choi:2006xb,Horton:2009ed,Badziak:2012yg,Lebedev:2005ge,Kaminska:2013mya} that one finds also `gaugino focus point' scenarios if a hierarchy among the gaugino masses is assumed at the GUT scale; for example such hierarchies can result from the breaking of a larger gauge group or from an underlying string theory \cite{Horton:2009ed,Kaminska:2013mya}.  

Before we turn to a detailed discussion of the FT in the MSSM assuming such a GUT model let us evaluate the effect of loop corrections on the inferred FT.

\subsection{Fine tuning including loop corrections}
\label{sec:loop}
The discussion of the impact of radiative corrections to the FT has a long history, see for instance the early works Refs.~\cite{deCarlos:1993rbr,Cassel:2009ps,Cassel:2010px}, where loop corrections were found to be potentially important. More recent papers however, such as Ref.~\cite{Kowalska:2014hza}, claim that the impact of loop corrections on the inferred FT are negligible, in particular for the FT with respect to $\mu$. 
To resolve this discrepancy consider the method proposed by Ref.~\cite{Baer:2012cf}  to incorporate the loop corrections by re-writing the loop corrected tadpoles as
\begin{equation}
\frac{M_Z^2}{2} \sim \mu^2 - m_{h_u}^2 - \Sigma_{uu}
\label{sigmauu}
\end{equation}
with $\Sigma_{uu} = \frac{\partial \Delta V}{\partial v^2}$ and
where all loop effects are absorbed into $\Sigma_{uu}$. This method was afterwards applied in many other works, see for instance Refs.~\cite{Baer:2012mv,Baer:2013vpa,Bae:2013qr,Baer:2013bba,Baer:2013vqa,Baer:2013ula,Baer:2013ssa,Baer:2013jla,Bae:2013hma,Han:2013kga,Baer:2013wza,Bae:2013bva,AbdusSalam:2013qba,Baer:2013faa,Baer:2013gva,Baer:2013xua,Baer:2013hoa,Bae:2014fsa,Baer:2014cua,Li:2014xqa,Baer:2014ica,Mustafayev:2014lqa,Bae:2014yta,Baer:2014kya,Bae:2015nva,Abe:2015xva,Bae:2015jea,Bae:2015rra,Tata:2015vwa,AbdusSalam:2015uba,Bae:2015efa,Altunkaynak:2015kia,Baer:2015tva,Li:2016ucz,Barger:2015xqk,Dimou:2015cmw,Bae:2016dxc,Calibbi:2016qwt,Baer:2016vls,Baer:2016usl,Cao:2016nix,Han:2016xet}. 
Crucially, when calculating the fine tuning, these papers all treat $\Sigma_{uu}$ as independent of $v$ and thus 
do not find a correlation between the tree and loop level terms when calculating $ \frac{\partial \ln
  v^{2}}{\partial \ln \mu}$. As a result the FT is found to be insensitive to loop corrections.
However in general  $\Sigma_{uu}$ does depend on the electroweak VEV $v$ (see below for an explicit example), implying immediately that this treatment is incorrect.

To estimate the effect of loop corrections on the FT we start with the loop corrected tadpole equation, working in the decoupling limit for simplicity (our numerical results
in later sections don't rely on this simplification)
\beq
0=(m_{h_u}^2+\mu^2+ \frac{1}{8}(g_1^2+g_2^2) v^2)v+\Sigma_u
\label{Tadpole}
\eeq
where $\Sigma_u={\partial\Delta V\over \partial v}$.
In order to take account of the correlations we  use the general parametrisation\footnote{In the following we assume that the $\Sigma_i$ are independent of $v$. This is not strictly speaking correct because of the appearance of $v$ in the logs in the loop corrections. This effect is neglected here in the analytical discussion but taken into account in our numerical checks. }
\begin{equation}
\Sigma_u = \Sigma_1 v + \Sigma_2 v^2 + \Sigma_3 v^3 \;.
\label{sigmau}
\end{equation}
Using eqs.~\eqref{Tadpole} and \eqref{sigmau} one may express the loop-level fine tuning in the form
\begin{equation}
\Delta_\mu = -\left(\frac{\partial\log M_Z^2}{\partial \log \mu} \right) \simeq \frac{8 \mu^2}{(g_1^2 + g_2^2 + 8 \Sigma_3) v^2 + 4 \Sigma_2 v}
\end{equation}
which reproduces the tree level result, eq.~\eqref{tree} in the limit  ($\Sigma_i=0$). 
The $\Sigma_i$ depend only very weakly on $\mu$ and for simplicity we have assumed $\frac{\partial \Sigma_i}{\partial \mu} = 0$ as in Ref.~\cite{Kowalska:2014hza}.
Thus, the dominant change in the FT measure does not come from the variation in $\Sigma$, but is due to the overall size of $\Sigma_{2,3}$.

\subsubsection{Example: stop corrections without mixing}
\begin{figure}[!t!]
\centering
\includegraphics[width=0.49\linewidth]{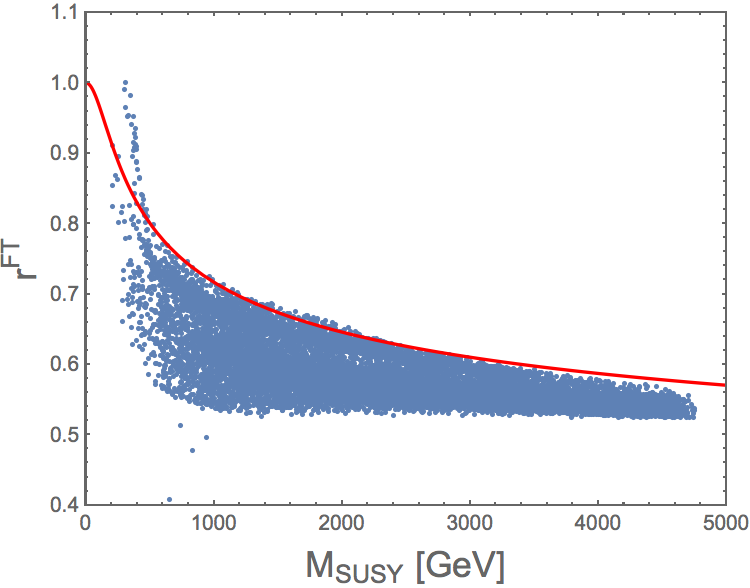}  
\includegraphics[width=0.49\linewidth]{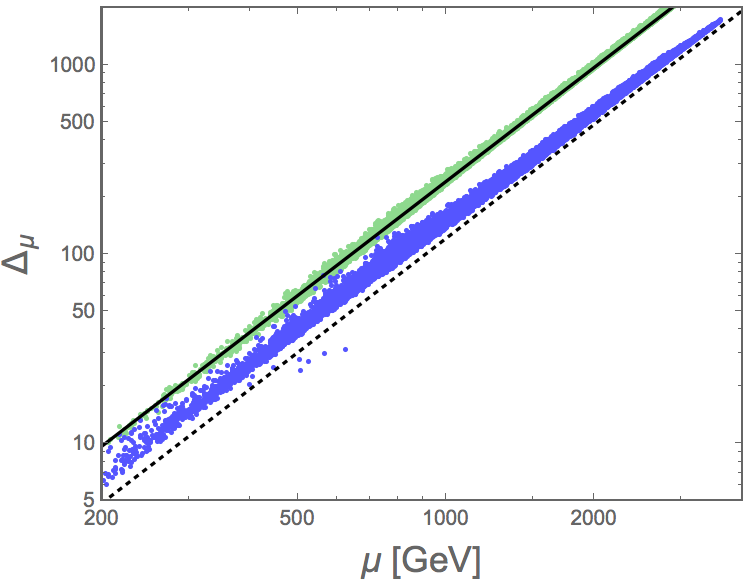} 
\caption{Left: Ratio of the fine tuning measures at loop- and tree-level as a function of the SUSY scale $M_\text{SUSY}$. 
The line indicates the analytical approximation as discussed in the text, while the points are from a numerical scan of the CMSSM.
Overall one can observe a reduction in FT of about 2 when going from tree- to loop-level. Right: Fine tuning with respect to $\mu$ at tree-level (lighter green) and loop-level
(darker blue) for the CMSSM case. The full (dashed) line correspond to the simple estimate of $\Delta^\text{tree}_\mu=2 \mu^2/M_Z^2$ ($\Delta^\text{loop}_\mu= \mu^2/M_Z^2$).}
\label{fig:0}
\end{figure} 
Let us discuss the simple example of stop corrections without mixing and with degenerate stop masses to illustrate our points.
The loop corrections to the scalar potential can be written as
\begin{equation}
\Delta V = \frac{3 \left\{\left(2 m_{\tilde t}^2+v^2 Y_t^2\right)^2 \left(2 \log \left(\frac{2 m_{\tilde t}^2+v^2 Y_t^2}{2 Q^2}\right)-3\right)+v^4 Y_t^4 \left(-2 \log \left(\frac{v^2 Y_t^2}{Q^2}\right)+3+\log (4)\right)\right\}}{256 \pi ^2} 
\end{equation}
where $m_{\tilde t}^2$ is a universal stop soft-breaking mass, $Y_t$ is the $\overline{\text{DR}}$ top Yukawa coupling and $Q$ is the renormalisation scale.
For this, one finds
\begin{equation}
\Sigma_{uu} \equiv \frac{\partial \Delta V}{\partial v^2}  = \frac{3m_{\tilde t}^2 Y_t^2 \left(\log \left(\frac{m_{\tilde t}^2}{Q^2}\right)-1\right)}{32 \pi ^2} + v^2 \frac{3 Y_t^4 \left(\log \left(\frac{2m_{\tilde t}^2+v^2 Y_t^2}{Q^2}\right)-\log \left(\frac{v^2 Y_t^2}{Q^2}\right)\right)}{64 \pi ^2} \;.
\end{equation}
Clearly $\Sigma_{uu}$ does depend on the electroweak vev $v$.
Using our parametrisation from above, the coefficients in eq.\eqref{sigmau} can be expressed as 
\begin{align}
 \Sigma_1 & = -\frac{3 m_{\tilde t}^2 Y_t^2 \left(-2 \log \left(\frac{2 m_{\tilde t}^2+v^2 Y_t^2}{Q^2}\right)+2+\log (4)\right)}{32 \pi ^2}\\
 \Sigma_2 & = 0 \\
 \Sigma_3 & = \frac{3 Y_t^4 \left(\log \left(\frac{2 m_{\tilde t}^2+v^2 Y_t^2}{Q^2}\right)-\log \left(\frac{v^2 Y_t^2}{Q^2}\right)\right)}{32 \pi ^2}
\end{align}
For this simple example, the change in the fine-tuning from tree- to loop-level can be approximated as follows
\begin{equation}
r^{FT} \equiv \frac{\Delta^{\rm Loop}}{\Delta^{\rm Tree}} = \frac{g_1^2+g_2^2}{g_1^2+g_2^2+8 \Sigma_3} \simeq \left(1 + \frac{3 Y_t^4 \log \left(\frac{ m_{\tilde t}^2}{m_t^2}\right)}{4 \pi ^2 (g_1^2+g_2^2)} \right)^{-1}
\end{equation}
where we set $Q^2=m_{\tilde t}^2$ and used $m_{\tilde t} \gg m_t$. In Fig.~\ref{fig:0} we plot $r^{FT}$ as function of the SUSY scale $M_\text{SUSY}$
showing that one can expect that the FT improves by about a factor of two if the loop corrections are correctly included. We also show the FT with respect to $\mu$
as a function of $\mu$ as well as the simple analytical estimate for the FT. Given that the reduction in FT is about a factor of 2 (see also ~\cite{Cassel:2009ps,Cassel:2010px} for a similar result), we propose to use 
\begin{equation}
\Delta^\text{loop}_\mu \simeq \frac{\mu^2}{M_Z^2}
\label{eq:muloop}
\end{equation}
when estimating the FT with respect to $\mu$.

\section{The MSSM with additional non-holomorphic soft-terms}
\label{sec:NHSSM}
\subsection{The new soft-terms}
In our discussion of the estimate of the FT we have seen that, despite the reduction in fine tuning due to loop effects, small FT prefers a moderately small $\mu$-term. Therefore the Higgsino mass $m_{\tilde H}$, which is mainly given by $\mu$ in the MSSM, is strongly limited by the requirement of small FT. However, this constraint can be softened if there are other sources of Higgsino mass. This leads us to consider the  possibility of additional soft SUSY breaking terms including a non-holomorphic contribution 
to the Higgsino mass,
\begin{align}
\mathcal{L}_{NH} = & \mu' \tilde h_d \tilde h_u \; + T'_{u,ij}  h_d^* \tilde u_{R,i}^* \tilde q_j +T'_{d,ij}  h_u^* \tilde d_{R,i}^* \tilde q_j + 
T'_{e,ij}  h_u^* \tilde e_{R,i}^* \tilde l_j  +  \text{h.c.}
 \label{newsofts}
\end{align}
In the context of the MSSM these terms are still `soft' in the sense that they do not lead to quadratic divergences at radiative order \cite{Girardello:1981wz}, such quadratic divergences only appearing in models with additional gauge singlet fields. Usually the non-holomorphic Higgsino mass term is simply not written down, as it can be reabsorbed into
the other soft-terms and superpotential parameters. As shown in \cite{Ross:2016pml} however, this term can be crucial when considering FT and should therefore not be absorbed.\footnote{The phenomenological impact of the soft-terms as well as the fine-tuning from the low-energy point of view has been studied recently in Ref.~\cite{Un:2014afa,Chattopadhyay:2016ivr}.}

\subsection{Impact and origin of large $\mu'$}
Of particular relevance to the Higgsino mass bound is the possibility of the Higgsino mass $\mu'$ in eq(\ref{newsofts}). While the soft-terms in eq.~(\ref{newsofts}) keep the 
minimum conditions in eqs.~(\ref{eq:tad1})--(\ref{eq:tad2}) unchanged, the $\mu'$ term shift the Higgsino mass to
\begin{equation}
m_{\tilde h} = \mu + \mu' 
\end{equation}

It has been pointed out in Ref.~\cite{Ross:2016pml} that a large $\mu'$ can have very important consequences on the FT. This is in particular the case when the constraints from the dark matter relic are included. The correct abundance of Higgsino dark matter is naturally obtained for Higgsino masses of around 1~TeV. However, this would immediately correspond to a FT of at least 120 as can be seen from eq.~\eqref{eq:muloop}. On the other hand, if the Higgsino mass stems to a large extent from $\mu'$, 1~TeV Higgsinos could still be consistent with a FT well below 50 \cite{Ross:2016pml}. We are going to extend the analyses of Ref.~\cite{Ross:2016pml} in sec.~\ref{sec:results}  by performing exhaustive parameter scans for a variety of different GUT boundary conditions. First, however, we discuss schemes in which a large  $\mu'$ term can arise in a UV complete model. 

\begin{figure}[tbp]
\vskip-0.1in
\begin{flushleft}
\hspace{3cm}
\includegraphics[width=8.9cm]{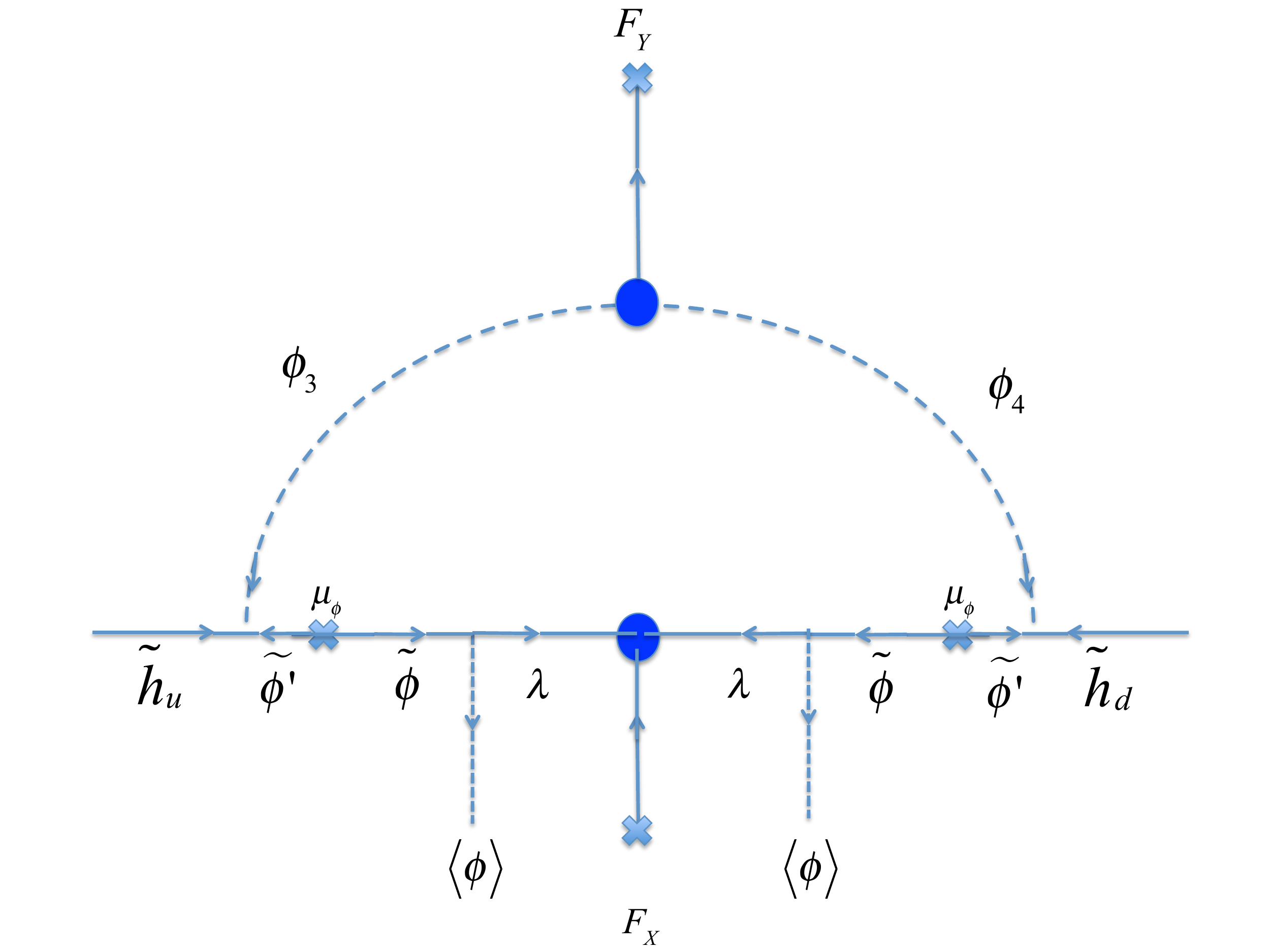} 
\end{flushleft}
\caption{{Graph generating Higgsino mass via gaugino mediation. }}
\label{figure_Higgsino}
\end{figure}

\begin{table}[t]
\label{tab:charges}
\begin{center}
  \begin{tabular}{ |c |c | c |c|}
   \hline
    Field&$SU(2)_L$ & $U(1)_Y$ &$U(1)'$ \\ \hline
  $ H_d$ & 2 & 1&0 \\ \hline
   $H_u$ & 2 & -1 &0\\ \hline 
$\Phi_3$ & 2 & -1&1 \\ \hline
$\Phi_4$ & 2 & 1&1 \\ \hline
$\Phi$ & 1& 0&1 \\ \hline
$\Phi'$ & 1 & 0&-1 \\ \hline
X & 1 & 0&0 \\ \hline
Y & 1 & 0&-2 \\ \hline
  \end{tabular}
  \caption{Transformation properties of fields under $SU(2)_L\times U(1)_Y\times U(1)'$}
  \label{tab:1}
\end{center}
\end{table}

As discussed by Martin \cite{Martin:1999hc}, in superspace co-ordinates the $\mu'$ term has the form 
\begin{equation}
\mu'\int {{d^4}\theta X{X^\dag }{{\overline D }_\alpha }\left( {H_d^\dag {e^V}} \right){{\overline D }^\alpha }{e^{ - V}}H_u^\dag } \;. 
\label{Martinop}
\end{equation}

In the MSSM the Higgsino gets mass via its coupling to the Majorana masses of the gauginos. For gaugino masses of $\mathcal{O}(M_\text{SUSY})$,  much greater than the EW breaking scale this mass is of $\mathcal{O}(v^2/M_\text{SUSY})$ and so is much smaller than the SUSY breaking scale. However this suggests that large Higgsino masses may be generated by coupling to heavy gauginos in the SUSY breaking sector.  As a concrete example consider a $U(1)'$ hidden sector with gaugino mass generated by the term $\frac{1}{M}\int {{d^2}\theta X{W^\alpha }{W_\alpha }} $ when $F_X\ne 0$.  In addition to the MSSM Higgs fields $H_{d,u}$ we add  Higgs portal mediator fields $\Phi_{3,4},\;\Phi,\;\Phi'$ and a further field $Y$ that also acquires an F-term. In Tab.~\ref{tab:1} we exhibit the transformation properties of the fields under the Standard Model gauge group $SU(2)_L\times U(1)_Y$ and the additional hidden sector $U(1)'$ gauge group. With the charge assignments shown the superpotential has the form 

\begin{equation}
W={\mu _\phi }\Phi \Phi ' + Y{\Phi _3}{\Phi _4} + \Phi '{\Phi _1}{\Phi _3} + \Phi '{\Phi _2}{\Phi _4}
\end{equation}
where $Y$ obtains a scalar VEV $\langle Y \rangle = M$ as well as an F-term VEV $F_Y$.
With these couplings the graph of Fig.~\ref{figure_Higgsino}  generates an  Higgsino mass 
$m_{\tilde{h_u}\tilde{h_d}} \propto \frac{{\mu _\phi ^2{{\left\langle \phi  \right\rangle }^2}{F_Y}{F_X}}}{{{M^7}}}$ where the messenger masses in the denominator depend on the details of the model. One may readily check that there is no equivalent one-loop scalar mass term for the MSSM Higgs fields.\footnote{
Another way to obtain an Higgsino mass without the SUSY related Higgs scalar mass term has been pointed out by Perez, Roy and Schmaltz \cite{Perez:2008ng}. It results from renormalisation group (RG) scaling from approximate strong dynamics in a hidden sector. The authors show that, starting with a supersymmetric $\mu$ term for the Higgs and an independent  soft Higgs scalar mass, the RG running drives the Higgs scale soft mass squared to $-\mu^2$, naturally cancelling the supersymmetric contribution to the Higgs mass squared term where the equality of coefficients is a natural result of the RG running. However, although this gives a Higgsino mass without a Higgs scalar mass,  this does not come from the operator in eq.~\eqref{Martinop}. As a result, even in models with Standard Model singlet superfields such as the NMSSM,  there are no quadratic divergences and so the mechanism can be applied to these models too. An analysis of this case is beyond the scope of this paper. 
A Higgsino mass without its scalar counterpart can also be generated via the Scherk-Schwarz mechanism (see \cite{Scherk:1979zr}). More recently constructions of natural SUSY models relying on the Scherk-Schwarz mechanism have received some interest \cite{Dimopoulos:2014aua,Garcia:2015sfa,Delgado:2016vib}. In this case the mass is comparable to the scale of compactification and the new states associated with the extra dimension spoil the simple gauge coupling unification considered in this paper (see however \cite{Huang:2016dtj} for an attempt to achieve gauge coupling unification also in these kinds of set-ups).}

\section{Numerical Results}
\label{sec:results}

\subsection{Setup, boundary conditions and constraints}
\label{sec:setup}

\subsubsection{Numerical setup}
For our analysis we have used the {\tt Mathematica} package {\tt SARAH} \cite{Staub:2008uz,Staub:2009bi,Staub:2010jh,Staub:2012pb,Staub:2013tta,Staub:2015kfa} in order to generate a spectrum generator based on {\tt SPheno} \cite{Porod:2003um,Porod:2011nf}. This provides a full-fledged spectrum generator including the RGEs for the standard SUSY terms up to two loops, based on Refs.~\cite{Martin:1993zk}, while the non-holomorphic soft-terms are fully included at one-loop level based on the results of Refs.~\cite{Jack:1999ud,Jack:1999fa}. The SUSY mass spectrum is calculated at full one-loop, while for the Higgs states the important two-loop corrections are covered in the effective potential approach \cite{Goodsell:2014bna,Goodsell:2015ira}: this includes also the one- and two-loop corrections to the Higgs mass via the non-holomorphic soft-terms. All important flavour constraints such as $b\to s\gamma$, $\Delta M_{B_q}$, $B\to l\bar l$ are checked via the {\tt FlavorKit} functionality \cite{Porod:2014xia}. \\
The routine to calculate the fine-tuning via {\tt SARAH}/{\tt SPheno} has been improved during this work to include also the loop corrections as discussed in sec.~\ref{sec:loop} \footnote{The changes are public with version 4.9.2 of {\tt SARAH}}. The numerical procedure works now as follows:
For each independent parameter (e.g. $m_0$, $m_{1/2}$, $A_0$, $\mu$, $B_\mu$) the following steps are performed for a given parameter point
\begin{enumerate}
 \item The numerical value of the considered parameter is varied at the GUT scale
 \item The two-loop RGEs are evaluated from the GUT to the SUSY scale
 \item The one-loop corrections to the tadpole equations are calculated using the initial VEVs
 \item The one-loop corrected tadpole equations are solved with respect to the VEVs
 \item The one-loop corrections to the tadpole equations are re-calculated using the found VEVs
 \item The last two steps are iterated, until the solution for the VEVs has converged
 \item The FT is calculated from the change in the VEVs or equivalently the $Z$ mass
\end{enumerate}
This iteration also ensures that  the VEV-dependence of all logs in the loop function is fully included. The overall fine-tuning is taken to be the maximal value of the fine-tuning calculated in this way, $\Delta = \text{max} \Delta_p$

\subsubsection{Boundary conditions}
In the following we numerically study the amount of FT in constrained versions of the MSSM with and without non-holomorphic Higgsino masses. 
We consider both the fully constrained versions where both the gaugino masses and all scalar masses unify at the GUT scale 
as well as the case of non-universal gaugino and Higgs masses. In all cases we assume universal squark and slepton masses
\begin{align}
& m_{\tilde q}^2 = m_{\tilde d}^2 = m_{\tilde u}^2 = m_{\tilde e}^2 = m_{\tilde l}^2 = {\bf 1} m_0^2 
\end{align}
and we parametrise the trilinear soft-terms as
\begin{equation}
T_i =  A_0 Y_i\,,\hskip1cm
T'_i = A_0' Y_i
\end{equation}
As usual we take the trilinear A terms to be proportional to the corresponding superpotential couplings and for simplicity we assume the
same relation for the $A_0'$ term. Our results are rather insensitive to this choice. 
The case of non-universal gaugino masses we parametrise via coefficients $a$ and $b$ at the GUT scale:
\begin{align}
M_1&=a \cdot m_{1/2} \\
M_2&=b \cdot m_{1/2}\\
M_3&=m_{1/2}
\end{align}
In total we study 6 different boundary conditions, which are defined in Tab.~\ref{tab:BC}

All parameters discussed above as well as the value of the Higgsino mass, $\mu'$, are defined at the GUT scale. In addition, we fix the values of $\mu$ and $B_\mu$ by requiring correct EW symmetry breaking (EWSB) and use $\tan\beta$ as well as the phase of $\mu$ as input. The fine-tuning is calculated with respect to
$(m_0,\, m_{1/2},\, A_0,\, A_0',\, \mu,\, \mu',\, B_\mu)$ and for the case of non-universal Higgs masses also with respect to $(m_{h_u}^2,\, m_{h_d}^2)$
We do not include any FT penalty for $a$ and $b$ assuming that they are fixed in the underlying theory\footnote{As discussed in e.g.~\cite{Kaminska:2013mya} there
are a number of high-scale scenarios which predict fixed ratios among the gaugino masses, in which case an infinitesimal variation of $a$ and $b$ is not meaningful.}. 

\begin{table}[hbt]
\begin{center}
\begin{tabular}{|c||c|c|c|c|c|c|c|}
\hline 
& $m_{h_u}^2$ & $m_{h_d}^2$ & $M_1$   &$M_2$   &$M_3$   & $\mu'$ & $A_0'$ \\
\hline
\hline 
CMSSM & $m_0^2$ & $m_0^2$ & $m_{1/2}$ & $m_{1/2}$ &$m_{1/2}$ & - & - \\
\hline 
MSSM-NUHM & $m_{h_u}^2$ & $m_{h_d}^2$ & $m_{1/2}$ & $m_{1/2}$ &$m_{1/2}$ & - & - \\
\hline 
MSSM-NUGM &  $m_0^2$ & $m_0^2$ & $a \cdot m_{1/2}$ & $b \cdot m_{1/2}$ & $m_{1/2}$ & - & - \\ 
\hline 
CNHSSM & $m_0^2$ & $m_0^2$ & $m_{1/2}$ & $m_{1/2}$ &$m_{1/2}$ & $\mu'$ & $A_0'$ \\
\hline 
NHSSM-NUHM & $m_{h_u}^2$ & $m_{h_d}^2$ & $m_{1/2}$ & $m_{1/2}$ &$m_{1/2}$ & $\mu'$ & $A_0'$ \\
\hline 
NHSSM-NUGM &  $m_0^2$ & $m_0^2$ & $a \cdot m_{1/2}$ & $b \cdot m_{1/2}$ & $m_{1/2}$ & $\mu'$ & $A_0'$ \\
\hline
\end{tabular}
\end{center}
\caption{Definition of the different boundary conditions used in this work.}
\label{tab:BC}
\end{table}

\subsubsection{Constraints}

A dedicated re-casting of the collider limits from SUSY searches is beyond the scope of this work. Therefore we require a simplistic cut on the gluino mass depending on the neutralino mass based on Ref.~\cite{ATLAS-CONF-2016-052,ATLAS-CONF-2016-078}.   
Of course, in realistic SUSY models, the limits on the different masses can be significantly weaker compared to the results of simplified models \cite{Aad:2015baa,Buckley:2016kvr}. 
In addition we require all charged electroweak states such as charginos to have masses larger than 100 GeV.
In addition, we require that the Higgs mass should be in the range $122 \, \text{GeV}<m_h<128 \, \text{GeV}$. Since the pseudoscalar and heavy Higgs are usually much heavier for all points which survive the cuts on the SUSY masses, the Higgs coupling rates are very SM-like and no additional constraints come from their measurements. 

In a second step we assume that the lightest neutralino makes up (part of) the dark matter and we assume standard thermal freeze out as the main production mechanism.
To have a consistent scenario we require that either the neutralino dark matter does not over-close the universe, i.e.\ $\Omega h^2 < 0.13$  (weak requirement), or that 
it explains the entire DM abundance in the universe\footnote{We allow for a somewhat larger range than the experimental error in the PLANCK results \cite{Ade:2015xua}, because the relic abundance is calculated at tree-level making the theoretical prediction somewhat more uncertain.}, $0.11 < \Omega h^2 < 0.13$ (strong requirement). We also require that the neutralino passes the latest direct detection limits from LUX \cite{Akerib:2016vxi}. To evaluate the LUX constraint we rescale the bound by $\Omega h^2$ in the case of an under-abundance. 

When discussing the dark matter properties of a given scenario, one should not forget that there is (at least) a gravitino in the particle spectrum beyond the usual 
SM superpartners, with a mass depending on the 
underlying supersymmetry breaking mechanism. If the gravitino is the LSP, it is a good dark matter candidate and the lightest neutralino 
becomes the NLSP and will decay into the gravitino. In this case the bounds from the relic abundance don't apply. If the gravitino is heavier, it can be an additional
source of dark matter production via late decays, depending on the cosmological evolution of the Universe. With this in mind it is obvious that the possible cuts
due to the dark matter properties are on a different footing than the limits from collider searches and hence we show the results of these cuts individually below.

\subsection{Results}
\subsubsection{The fully constrained MSSM}
\label{sec:CMSSM}
\begin{figure}
\includegraphics[width=0.49\linewidth]{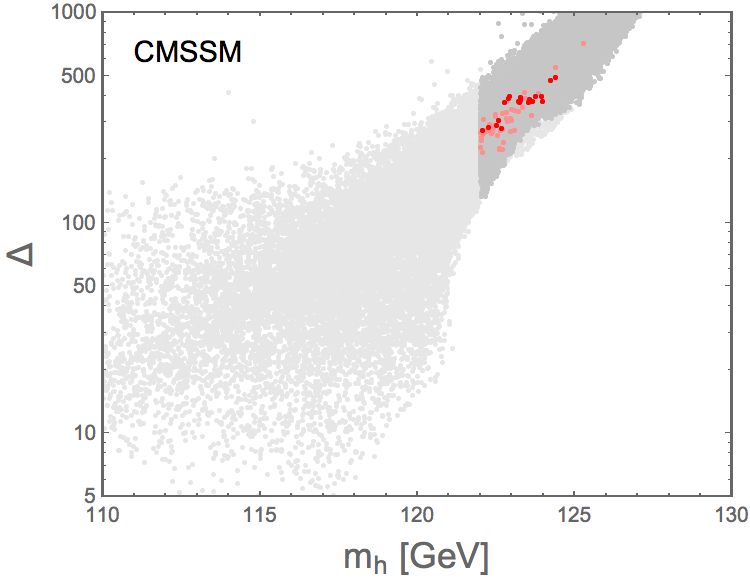}\hfill 
\includegraphics[width=0.49\linewidth]{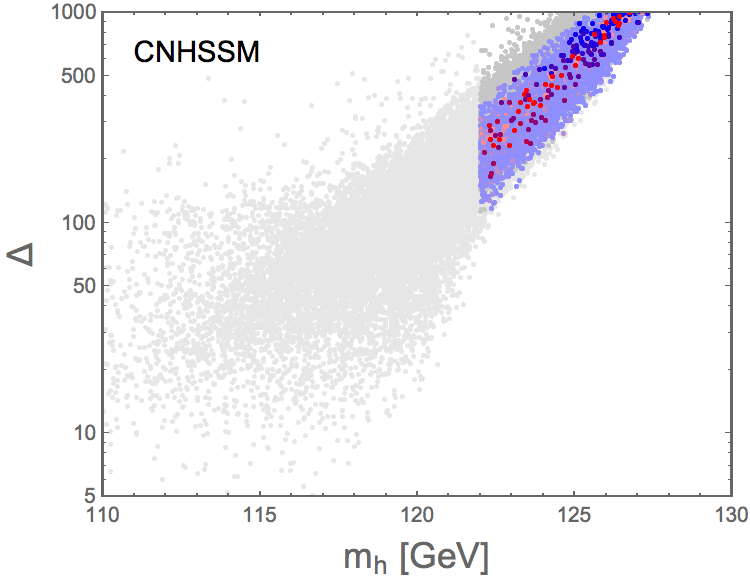}
\includegraphics[width=0.49\linewidth]{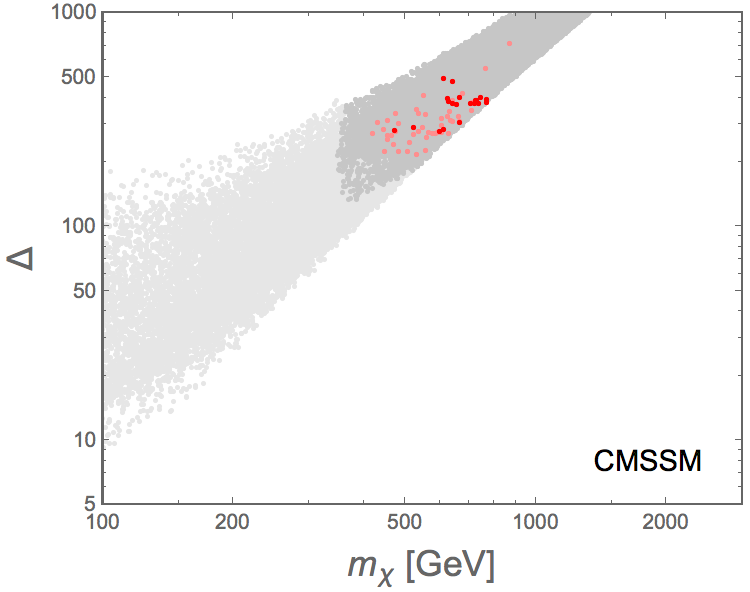}\hfill 
\includegraphics[width=0.49\linewidth]{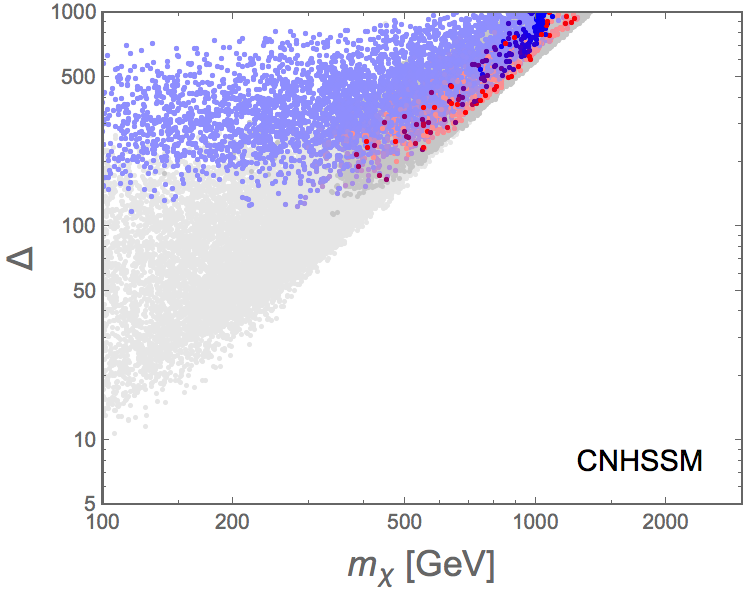}
\caption{Top: The FT vs.\ the mass of the MSSM-like Higgs in the CMSSM (left) and the CNHSSM (right).  Light gray points are before any cuts. Medium gray points are after bounds from collider searches have been taken into account. The light coloured points in addition pass the weak dark matter cuts while the dark coloured points include the lower bound on $\Omega h^2$ and hence require the correct relic abundance. It can be seen that the lightest neutralino in the CMSSM is always a bino, while in the CNHSSM also a Higgsino LSP is possible (see text for colour coding).
Bottom: FT vs.\ the mass of the lightest neutralino.} 
\label{fig:CMSSM_mh}
\end{figure}
\begin{figure}
\includegraphics[width=0.49\linewidth]{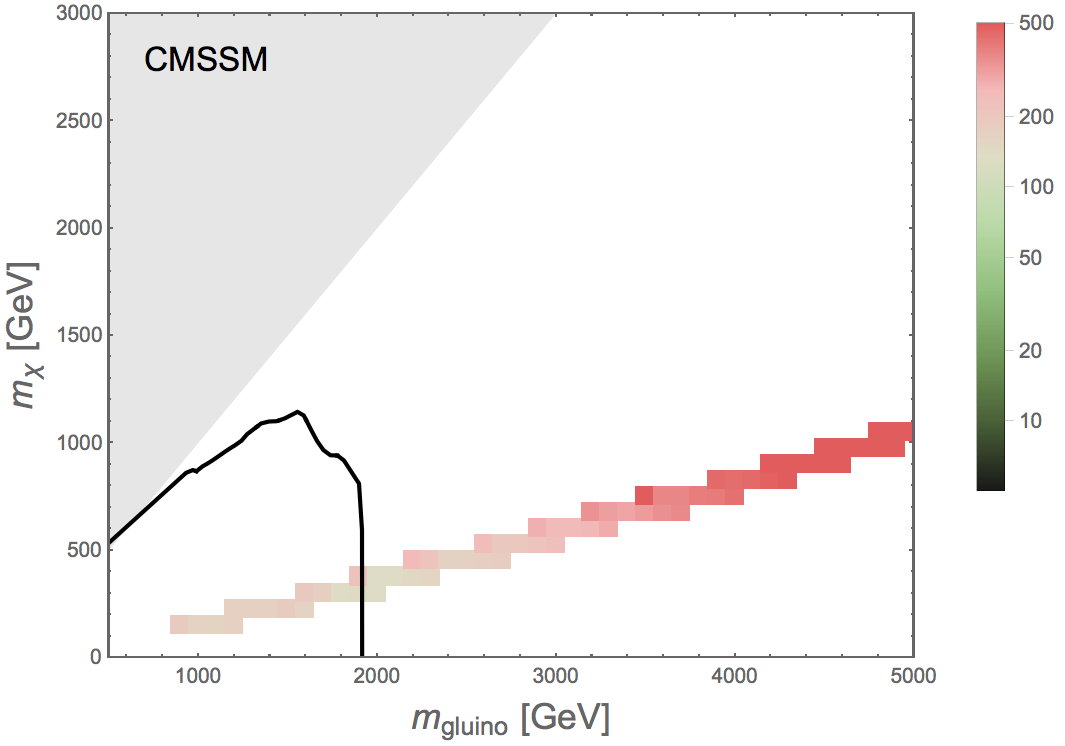}\hfill 
\includegraphics[width=0.49\linewidth]{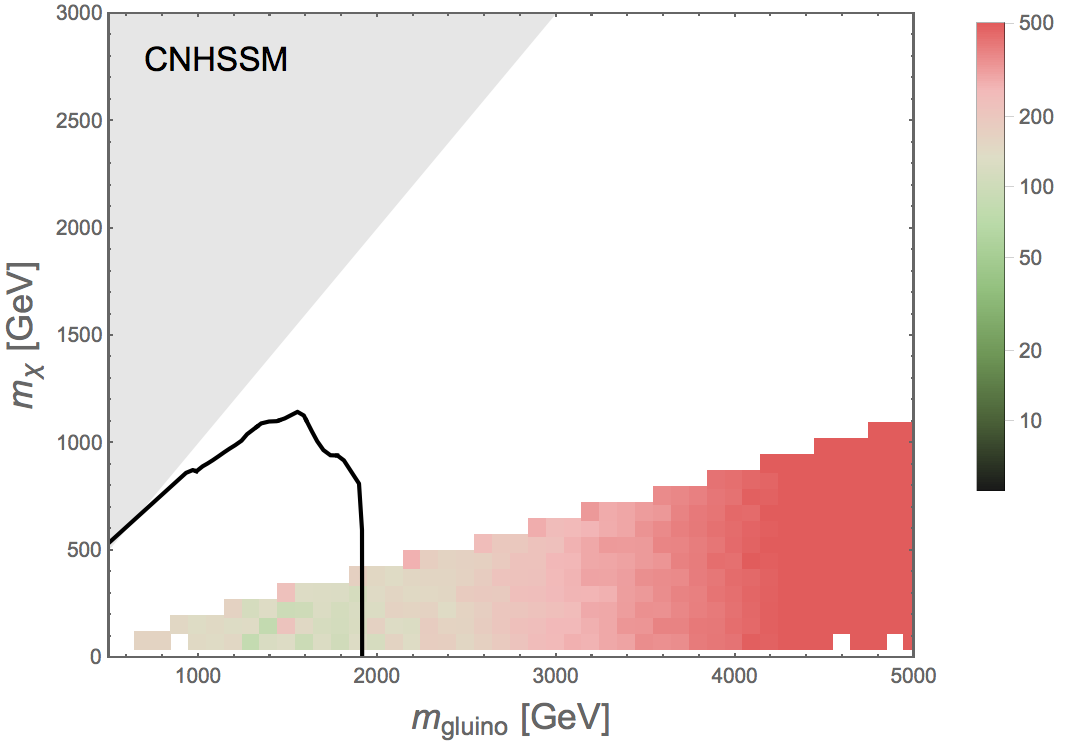}
\includegraphics[width=0.49\linewidth]{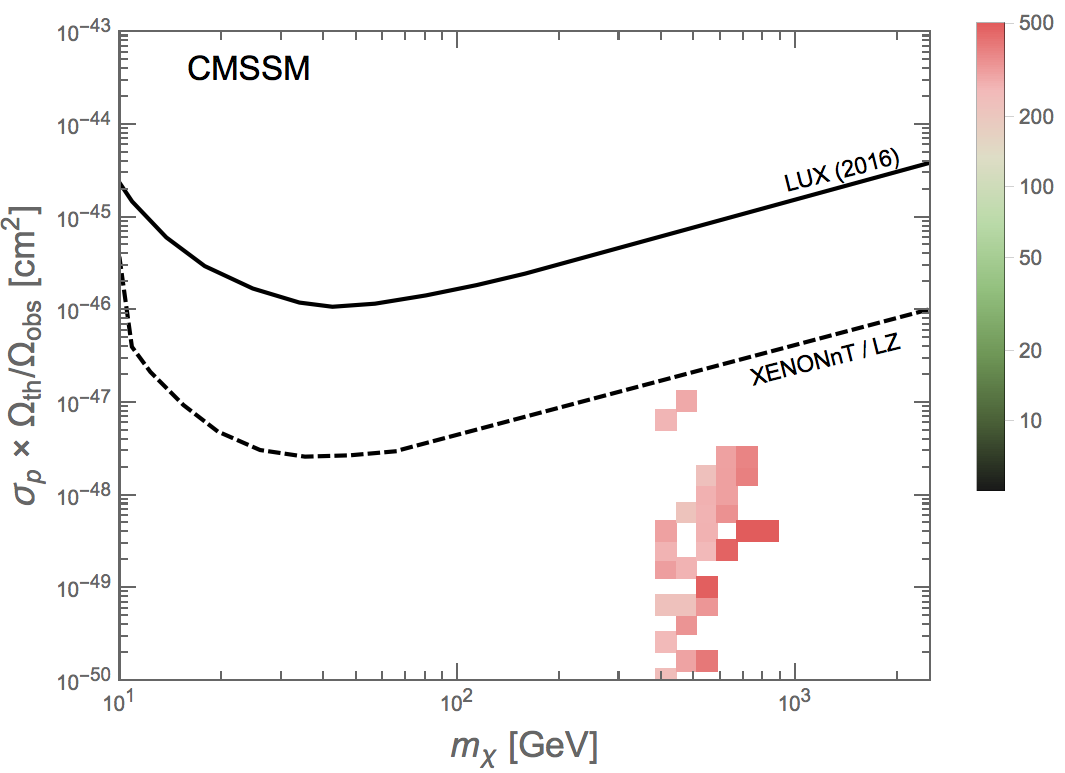}\hfill 
\includegraphics[width=0.49\linewidth]{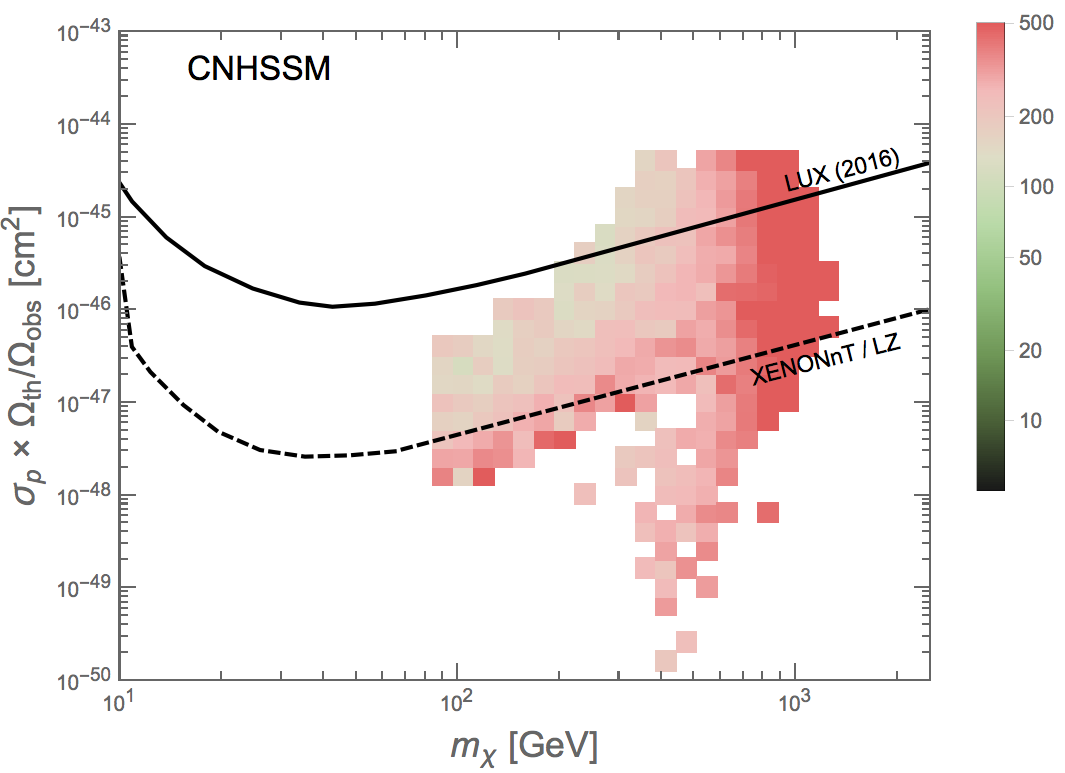}
\caption{Top: The FT in the gluino-LSP plane for the CMSSM (left) and the CNHSSM (right). The Higgs mass bounds as well as collider constraints are applied except the
cut on the gluino mass which is shown explicitly by the black line. Bottom: FT in the plane relevant for dark matter direct detection, where in addition to the collider constraints the 
upper bound on the neutralino relic abundance has been applied. The current LUX bound as well as the sensitivity of future experiments is also indicated by the black lines.} 
\label{fig:CMSSM_gluino}
\end{figure}
Let us start our discussion of the fine-tuning in the fully constrained MSSM with (CNHSSM) and without (CMSSM) the new soft terms.
Although FT in the CMSSM has been widely discussed, we redo the analysis here to take account of the loop corrections discussed above that have often been missed. 
Specifically we perform a random scan in the following parameter space:
\begin{eqnarray*}
& 0 <  m_0 < 3000~\text{GeV} & \\
& 100 <  m_{1/2} < 3000~\text{GeV} & \\ 
& 10 <  \tan\beta < 50 & \\
& -3000~\text{GeV} <  A_0 < 3000~\text{GeV} & \\
& \text{sign} \, \mu \pm 1 &
\end{eqnarray*}
For the CNHSSM we in addition scan over
\begin{eqnarray*}
& -2000~\text{GeV} < \mu' < 2000~\text{GeV} & \\
& -3000~\text{GeV} < A_0' < 3000~\text{GeV} & 
\end{eqnarray*}
The main results of these scans are shown in Figs.~\ref{fig:CMSSM_mh} and \ref{fig:CMSSM_gluino}.
In Fig.~\ref{fig:CMSSM_mh} we show the fine tuning against the SM-like Higgs mass and against
the lightest neutralino mass for the CMSSM and the CNHSSM.
Here and in the following we distinguish four cases:
\begin{itemize}
 \item Points with a well defined spectrum without any further cuts (light gray)
 \item Points passing the cuts on the Higgs mass and LHC limits (medium gray)
 \item Points passing the weak dark matter cuts (medium colour)
 \item Points passing the strong dark matter cuts (dark colour)
\end{itemize}
The coloured points contain additional information about the composition of the lightest neutralino, with red for a bino, green for a Wino and blue for a Higgsino.
Mixtures between these different states are shown via RGB colour coding. 

As expected from the discussion in sec.~\ref{sec:loopFT}, compared to a tree-level analysis, the FT becomes smaller by about a factor of 2 if the loops are included properly.
If only the collider bounds are applied, the smallest FT we find within the CMSSM is about 130 (cf.~Tab.~\ref{tab:FT} for a summary of our findings for all considered cases)
and quickly rises with the Higgs and neutralino masses.
For a unified gaugino mass $m_{1/2}$ at the high scale it is well known that at the low scale the ratios of gaugino mass parameters roughly scale as $M_3:M_2:M_1 \sim 6:2:1$.
Hence the bino is always the lightest gaugino for a universal $m_{1/2}$. In addition Higgsinos turn out to be typically significantly heavier, such that the lightest neutralino
is an almost pure bino, which can also be inferred from the colour coding in Fig.~\ref{fig:CMSSM_mh}.
Due to its rather small couplings a bino LSP has a small annihilation cross section and is therefore typically overproduced implying that only a small number of points pass all the cuts. 
This is because one needs to sit in fine-tuned co-annihilation or resonance regions, which is a potential tuning beyond the one in the electroweak sector. 

For the case of the CNHSSM, the picture after collider constraints have been taken into account is rather similar as far as the FT is concerned, with the smallest FT
of about 110. The large difference however concerns the composition of the lightest neutralino: Due to the extra contribution to the Higgsino mass light Higgsinos
are well possible, and it is much easier to obtain the correct relic abundance, as can be seen from Fig.~\ref{fig:CMSSM_mh}. However, only the usual $\mu$-term enters
the electroweak symmetry breaking conditions and hence there is no difference in this respect compared to the CMSSM: the usual $\mu$-term is still large and while the
extra term $\mu'$ can make the Higgsinos much lighter, the FT is still sizeable.

\subsubsection*{Future prospects}
In the coming years the LHC will continue to collect large amounts of data and simultaneously dark matter direct detection experiments will improve
the bounds on the dark matter nucleon scattering cross section by about 2 orders of magnitude. Given these expected improvements it is interesting to ask
how naturalness will be constrained in a given scenario in the foreseeable future. The answer can be inferred from Fig.~\ref{fig:CMSSM_gluino}. In the top panel 
we show the FT in the $m_\chi  - m_{\tilde{g}}$ plane to estimate the sensitivity of future collider searches. 
In the bottom panel we show the FT in the $m_\chi  - \sigma_p$ plane, where $\sigma_p$ is the DM nucleon scattering cross section. 
Here we rescale the DM nucleon scattering cross section for the case of an LSP underabundance in order to be able to sensibly compare to direct detection bounds,
$\sigma_p \rightarrow \sigma_p \cdot \Omega_\text{DM}^{\text{th}} / \Omega_\text{DM}^\text{obs}$.
For both cases we show the current bounds in the given plane, with the other collider bounds taken into
account in both cases and in addition the dark matter cuts applied to the bottom panels. In the bottom panels we also show the expected sensitivity of the planned XENONnT
or LZ \cite{Akerib:2015cja} experiments. We see that for the CMSSM the valid points in the $m_\chi  - m_{\tilde{g}}$ plane are
on a line with $m_\chi \sim 1/6 m_{\tilde{g}}$ as expected for a bino LSP due to the gaugino mass relations. 
For the CNHSSM the LSP can be lighter than $1/6 m_{\tilde{g}}$ if there is a sizeable Higgsino component, but of course not heavier (a heavier Higgsino would no longer be the LSP). In both cases the FT significantly increases with increasing gluino mass, with a minimal FT for $m_{\tilde{g}} \ge 3$ TeV ($\ge 5$ TeV)
of about 230 (700) in both cases (cf.~Tab.~\ref{tab:FT}). 
Constraining naturalness even more may be a more precise (theoretical) determination of the Higgs mass, given the steep slope in FT. 
Direct detection experiments are typically not sensitive with respect to the FT within the CMSSM, due to the small couplings of the bino LSP, while for the
CNHSSM the remaining parts of parameter space with FT about 100 will be tested by future direct detection experiments.

\subsubsection{Non-universal Higgs masses}
\begin{figure}
\includegraphics[width=0.49\linewidth]{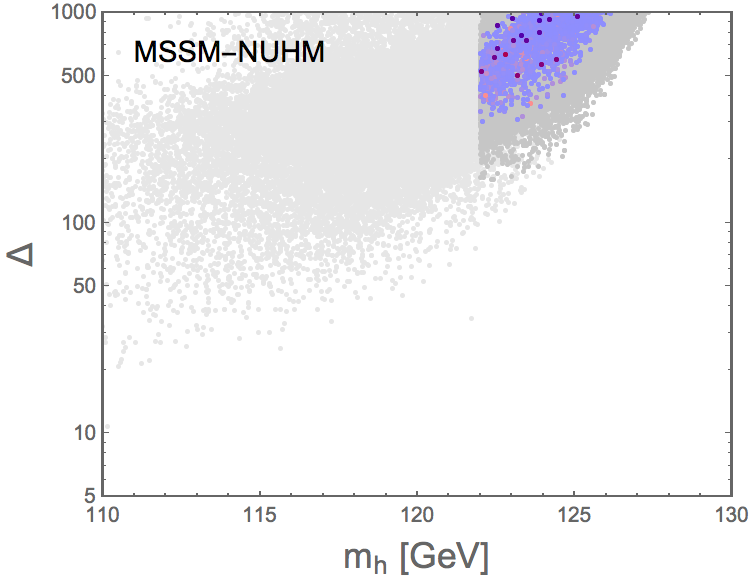}\hfill 
\includegraphics[width=0.49\linewidth]{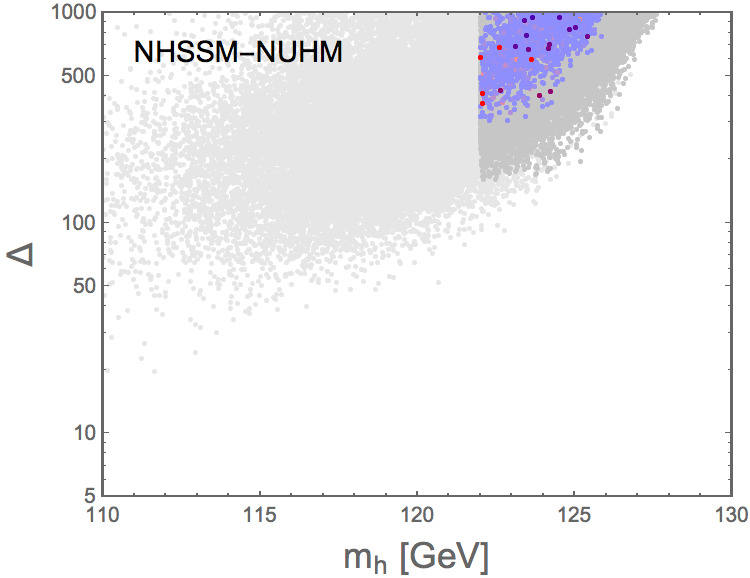}
\includegraphics[width=0.49\linewidth]{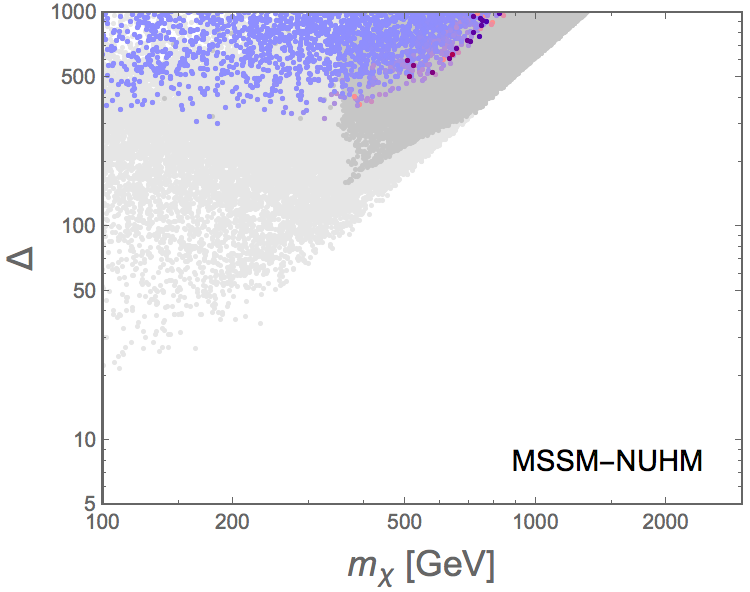}\hfill 
\includegraphics[width=0.49\linewidth]{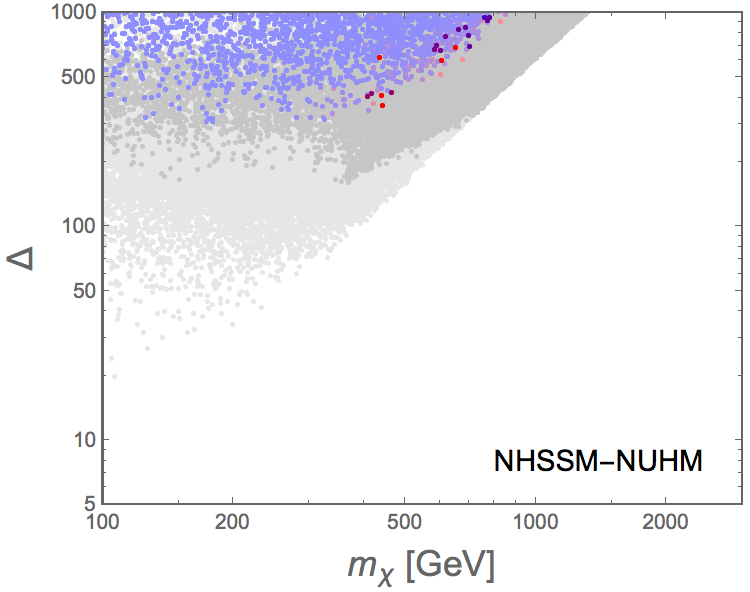}
\caption{Same as Fig.~\ref{fig:CMSSM_mh} but for the case of non-universal Higgs masses.} 
\label{fig:MSSMNUHM_mh}
\end{figure}
\begin{figure}
\includegraphics[width=0.49\linewidth]{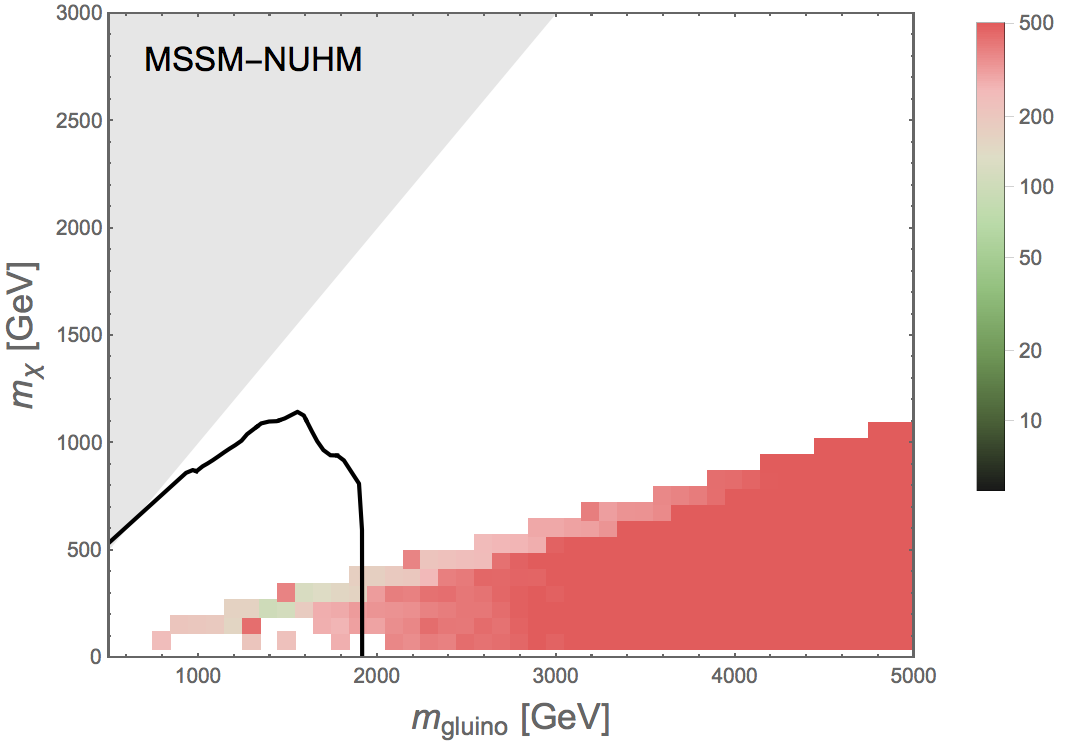}\hfill 
\includegraphics[width=0.49\linewidth]{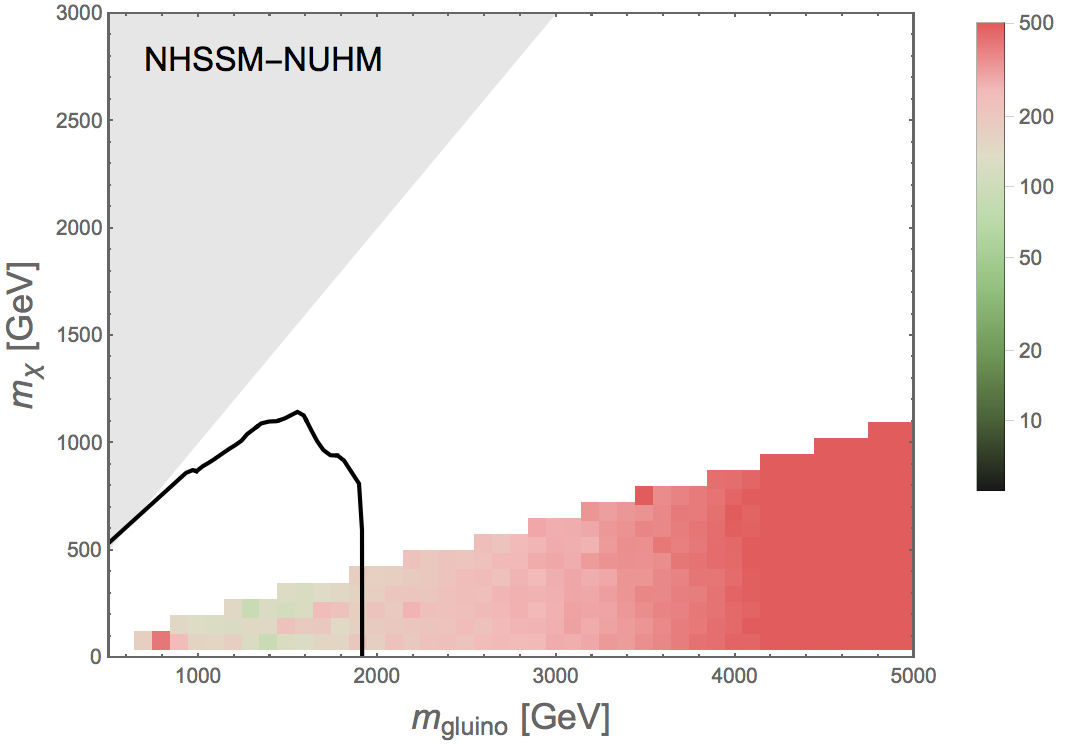}
\includegraphics[width=0.49\linewidth]{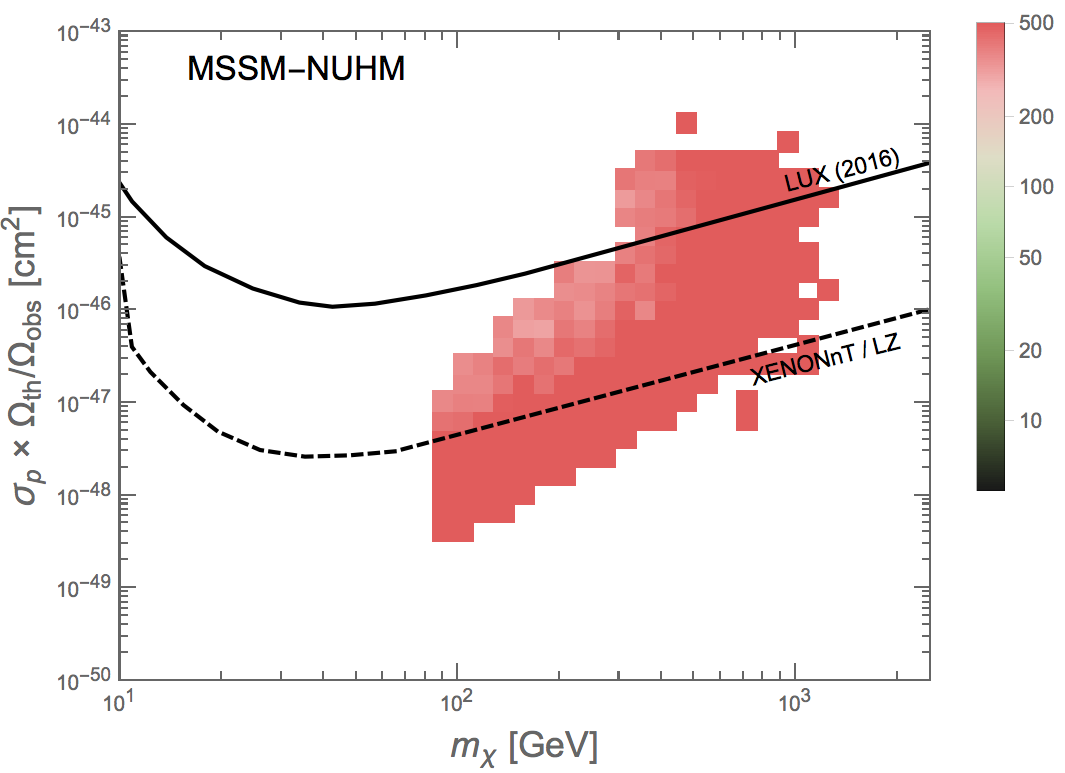}\hfill 
\includegraphics[width=0.49\linewidth]{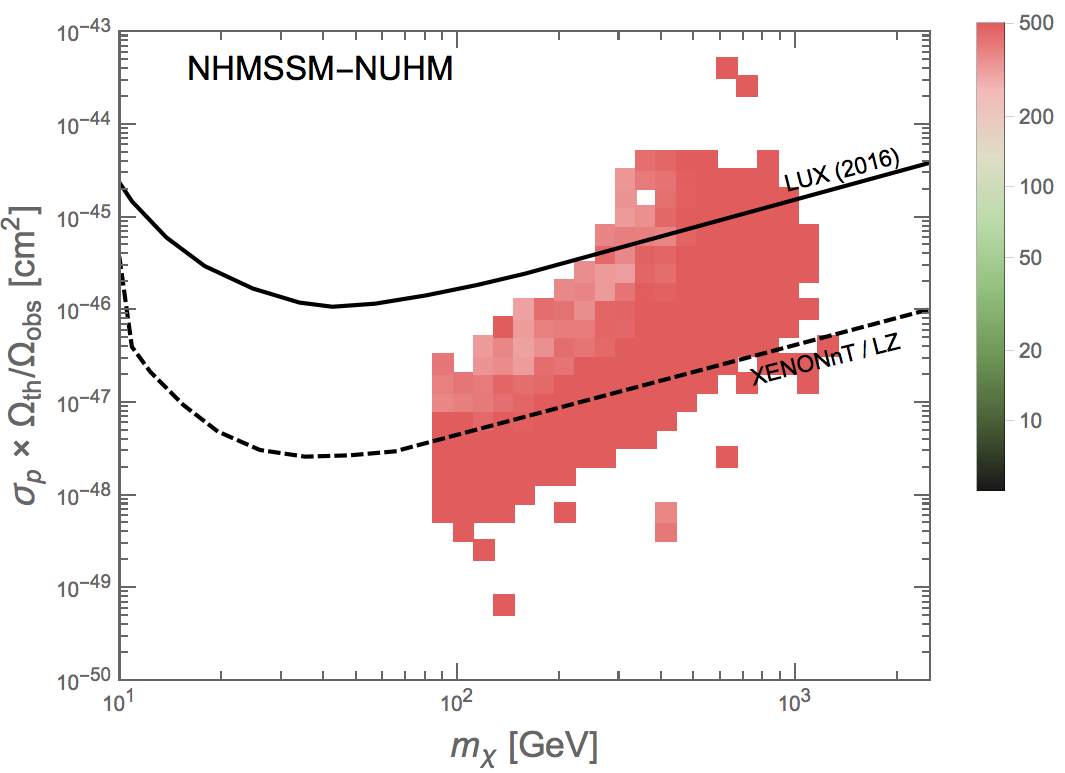}
\caption{Same as Fig.~\ref{fig:CMSSM_gluino} but for the case of non-universal Higgs masses.} 
\label{fig:MSSMNUHM_gluino}
\end{figure}
Non-universal Higgs masses are often discussed as a first step beyond the CMSSM. One of the motivations is that in many GUT models different masses
for the Higgs fields are possible, as they usually originate from a different GUT multiplet compared to the matter fields.
We therefore briefly discuss the results for the MSSM with non-universal Higgs mass terms with and without the new soft terms (denoted NHSSM-NUHM and MSSM-NUHM
respectively). In order to minimise bias in our results we solve the tadpole equations not only with respect to $\mu$ and $B\mu$ as before but also
with respect to the soft Higgs masses $m_{h_d}^2$ and $m_{h_u}^2$ and overlay the different scans.
Similarly as before our results are summarised in Figs.~\ref{fig:MSSMNUHM_mh} and \ref{fig:MSSMNUHM_gluino}.
It turns out that after collider cuts have been taken into account the resulting FT is even slightly worse than in the fully constrained versions.
This is likely due to the fact that the approximate focus point behaviour of a universal $m_0$ is no longer present. On the other hand the LSP
is typically a mixture of bino and Higgsino and a number of points have the correct relic abundance. Regarding future prospects this scenario is again
rather similar to the previous case, with FT increasing significantly with gluino mass.

\subsubsection{Non-universal gaugino masses}
\begin{figure}[hbt]
\begin{center}
\includegraphics[width=0.49\linewidth]{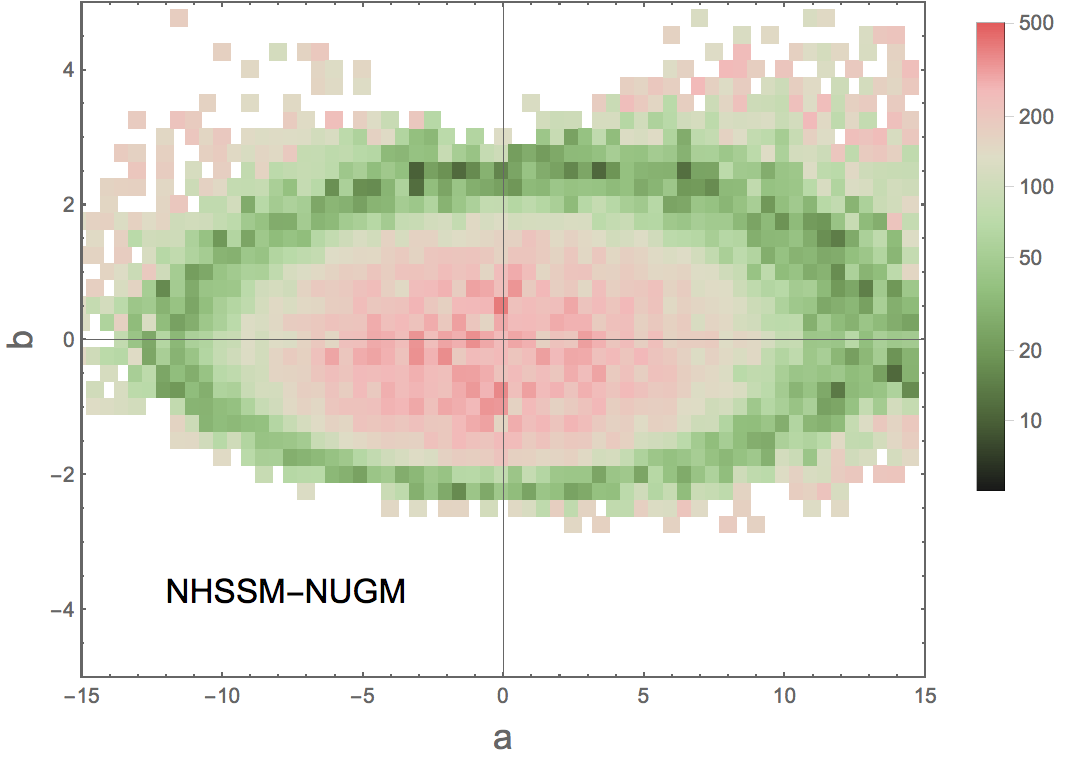} 
\end{center}
\caption{Fine tuning in the (a,b) plane after the LHC and Higgs cuts, showing that the gaugino focus point is realised in an ellipsoid in the (a,b) plane.}
\label{fig:MSSMNUGM_ab}
\end{figure}
\begin{figure}
\includegraphics[width=0.49\linewidth]{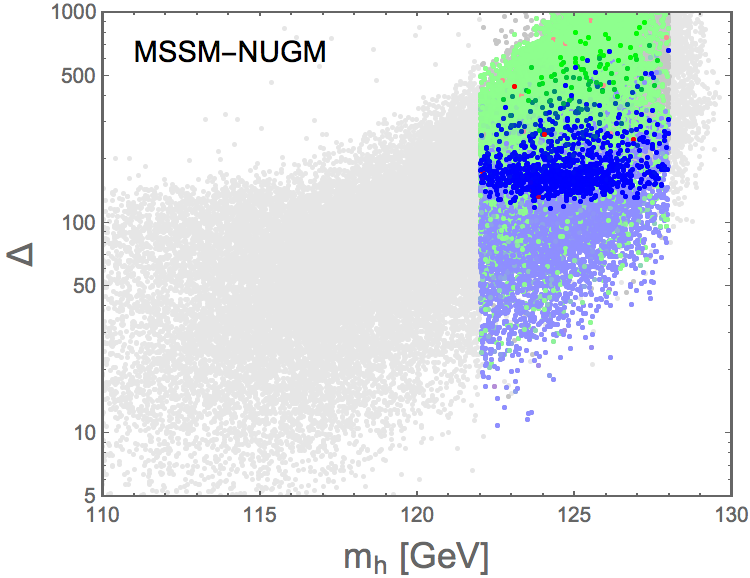}\hfill 
\includegraphics[width=0.49\linewidth]{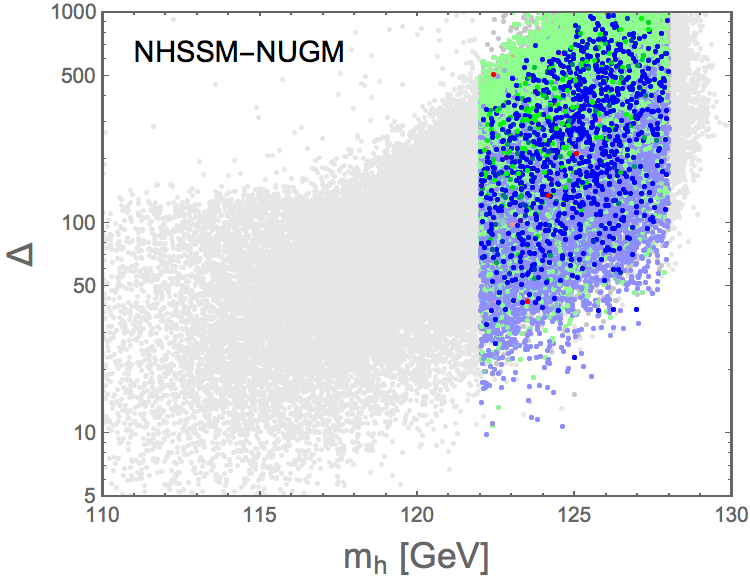}
\includegraphics[width=0.49\linewidth]{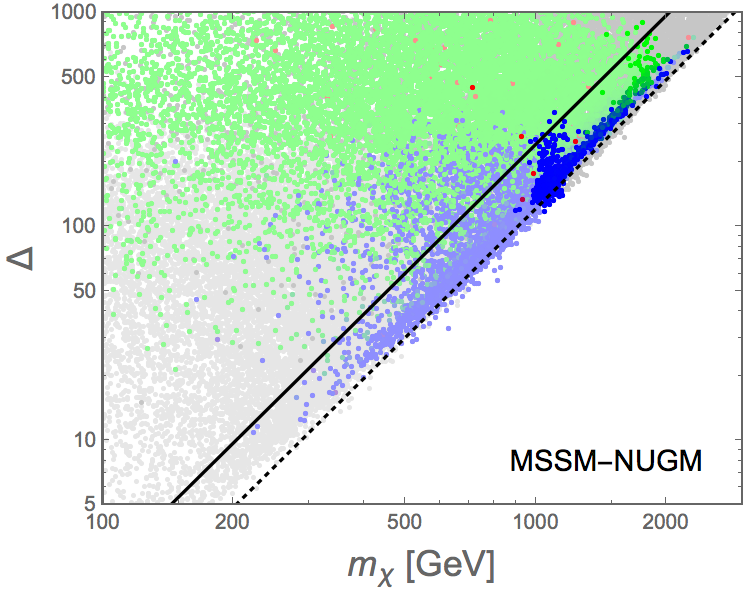}\hfill 
\includegraphics[width=0.49\linewidth]{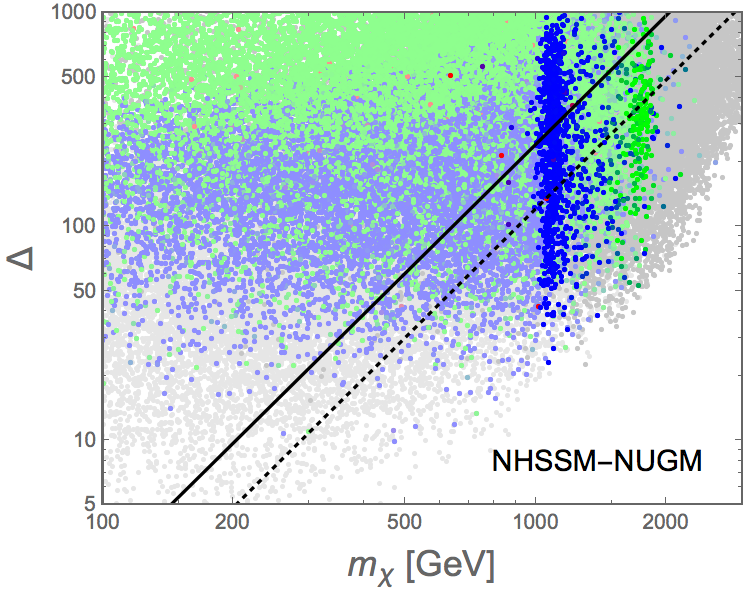}
\caption{Same as Fig.~\ref{fig:CMSSM_mh} but for the case of non-universal gaugino masses. In addition we indicate the smallest FT expected for the case of a Higgsino LSP if
its mass is due to the usual $\mu$ term only. The full (dashed) line correspond to the simple estimate of $\Delta^\text{tree}_\mu=2 \mu^2/M_Z^2$ ($\Delta^\text{loop}_\mu= \mu^2/M_Z^2$).} 
\label{fig:MSSMNUGM_mh}
\end{figure}
\begin{figure}[hbt]
\includegraphics[width=0.49\linewidth]{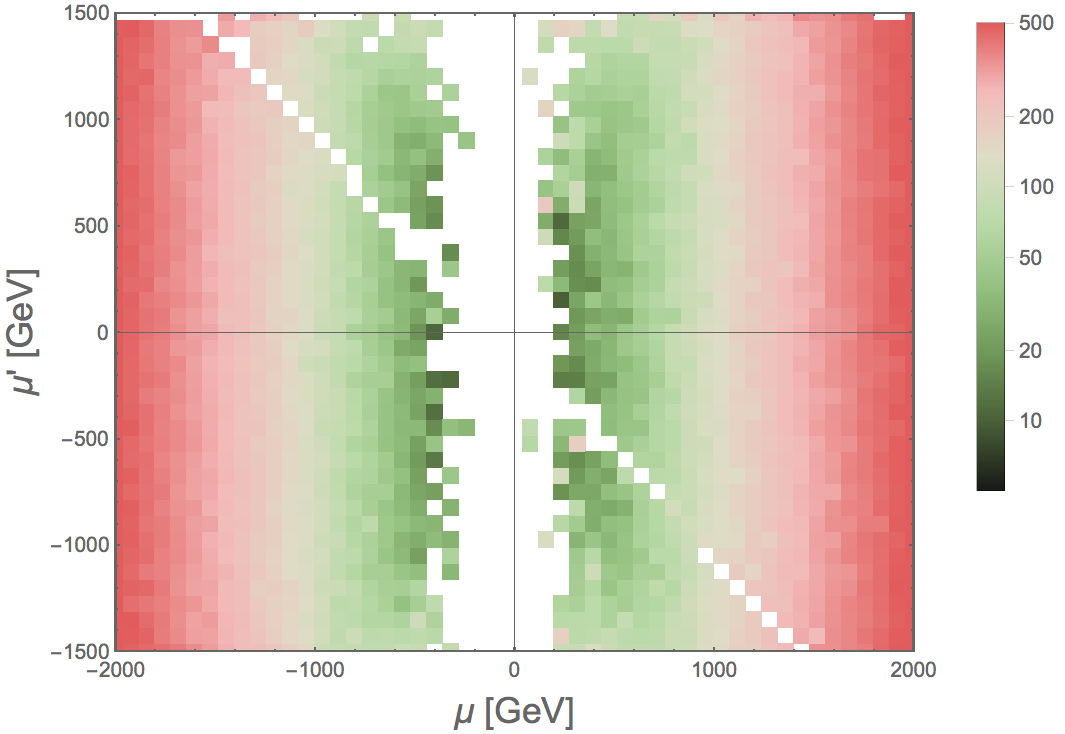} \hfill
\includegraphics[width=0.49\linewidth]{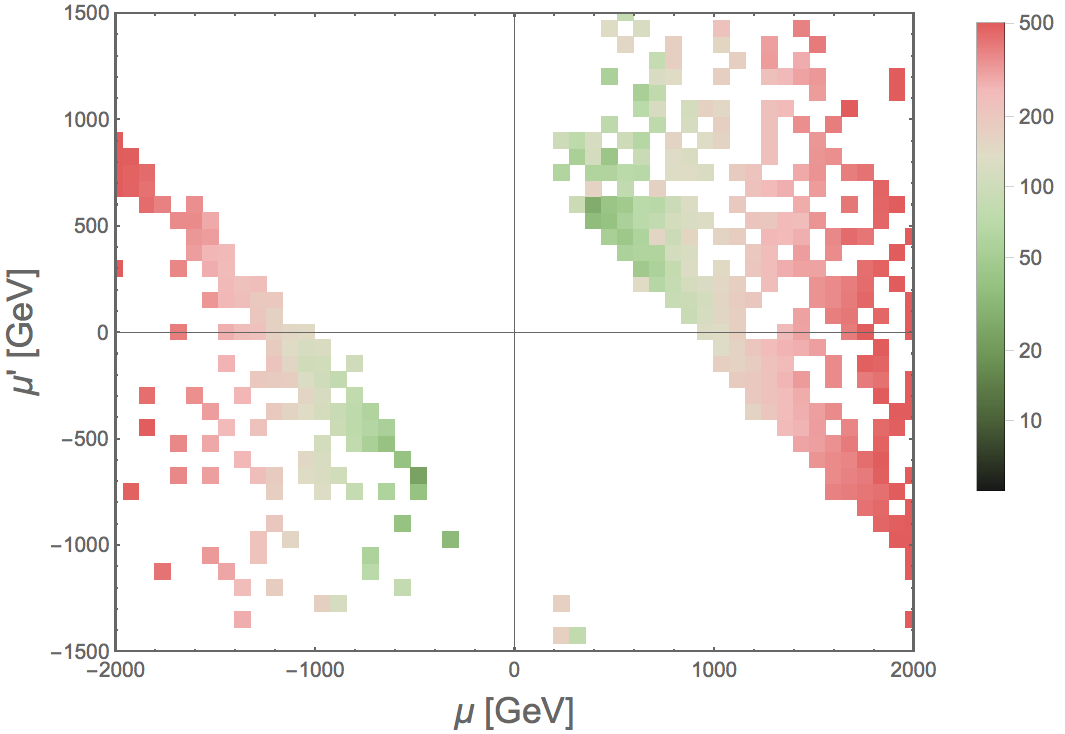} 
\caption{Fine tuning in the $\mu - \mu'$ plane, largely determining the Higgsino mass with collider constraints (left) and with additional strong DM constraints (right).}
\label{fig:FTmumup}
\end{figure}
Let us finally come to the case of non-universal gaugino masses, i.e.\ we relax the assumption that all gaugino masses unify at the GUT scale 
and allow for the following ratios in the gaugino masses at the high scale:
\begin{eqnarray*}
&-15 < a=\frac{M_1}{M_3} < 15 &\\
&-5 < b=\frac{M_2}{M_3} < 5 & 
\end{eqnarray*}
Non-universal gaugino masses have attracted quite some attention due to the existence of a gaugino focus point, allowing for a significant reduction in FT. 
This is illustrated in  Fig.~\ref{fig:MSSMNUGM_ab} where we show the FT as function of the gaugino mass ratios $a$, $b$.
Note that these ratios can still be consistent with an underlying GUT structure, see e.g.~\cite{Horton:2009ed} for more details.

Similarly as before the main results are summarised in Figs.~\ref{fig:MSSMNUGM_mh} and \ref{fig:MSSMNUGM_gluino}.
We see that for both cases, the MSSM-NUGM as well as the NHSSM-NUGM, a large reduction in FT is possible. Also the increase of FT with $m_h$ is significantly
milder than in the previous cases. After the collider constraints are taken into account the FT in both cases can be as small as 10.
From Fig.~\ref{fig:MSSMNUGM_mh} also the flexibility within the gaugino sector is evident, allowing for bino (red), Higgsino (blue) and Wino (green) LSPs. 
Higgsinos as well as Winos annihilate rather efficiently in the early Universe, leading to an underabundance in the DM relic density if their masses are below a TeV.
Therefore requiring the weak dark matter cut with no lower bound on the relic abundance, the FT stays as small as 10 in both cases (requiring however an alternative DM source such as axions or axinos or a different cosmological history).  
When applying the strong DM cut to explain the dark matter via the neutralino LSP, the Higgsino (Wino) mass gets pushed to about 1 TeV (2 TeV) as can be seen in
Fig.~\ref{fig:MSSMNUGM_mh}.\footnote{Note that it is beyond the scope of this work to include Sommerfeld enhancement effects which would slightly shift the mass of
the Wino to acquire the correct relic abundance.} In the MSSM this requires a $\mu$ term of about 1 TeV, resulting in significant tuning (of about 120).
In the NHSSM large Higgsino masses can be obtained due to the extra $\mu'$ contribution without a large FT penalty, allowing for FT of about 20 with the
correct relic abundance.
For clarity we show black lines in Fig.~\ref{fig:MSSMNUGM_mh} to illustrate this. The full black line corresponds to the expectation $\Delta^{\rm tree} = 2 \frac{M_\chi^2}{m_Z^2}$ from the MSSM where the Higgsino mass is up to small corrections given by the $\mu$-term. The dotted black line is a rough estimate for the loop corrections using $\Delta^{\rm loop} = \frac{1}{2} \Delta^{\rm tree}$. One can see that this approximation works well in the usual MSSM while in the presence of an extra contribution to the Higgsino mass the FT can be much smaller, since the Higgsino mass and $\mu$ term are now independent.
To illustrate this further, we show in Fig.~\ref{fig:FTmumup} the minimal FT in the ($\mu$,$\mu'$)-plane after the collider cuts (left) and after the additional requirement of the correct relic abundance (right). It can be seen that before the lower bound on the relic abundance is imposed there is no preference for $\mu' \neq 0$,
which is however strongly preferred once this bound is implemented.

\subsubsection{Future prospects}
\begin{figure}
\includegraphics[width=0.49\linewidth]{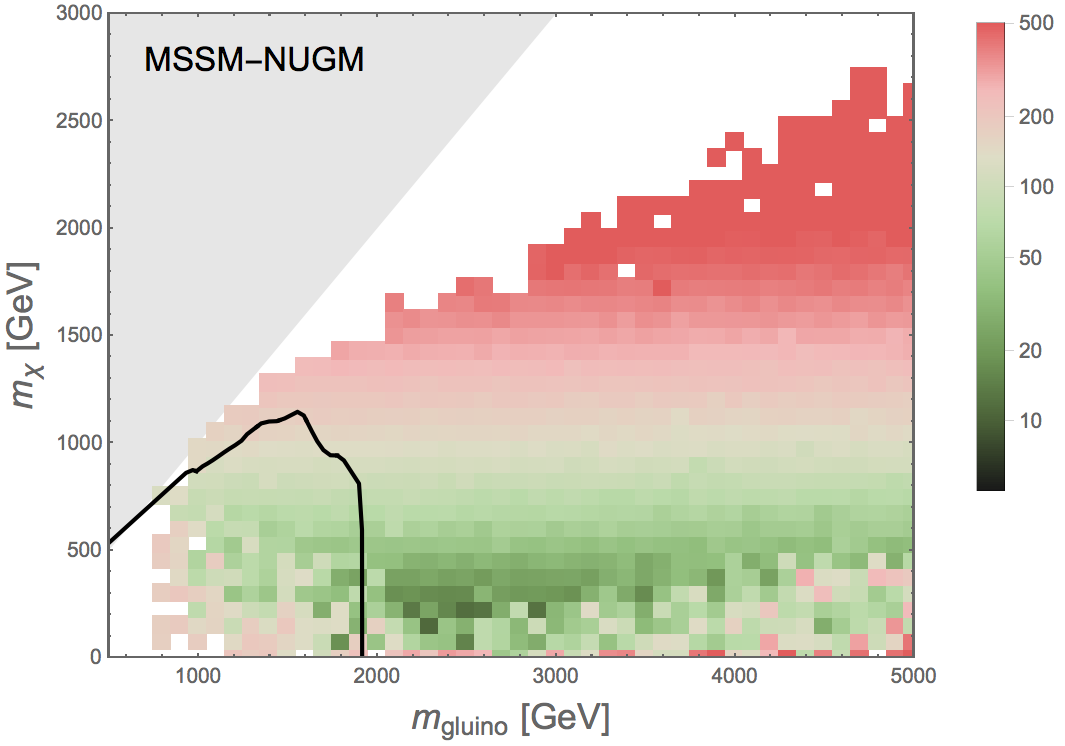}\hfill 
\includegraphics[width=0.49\linewidth]{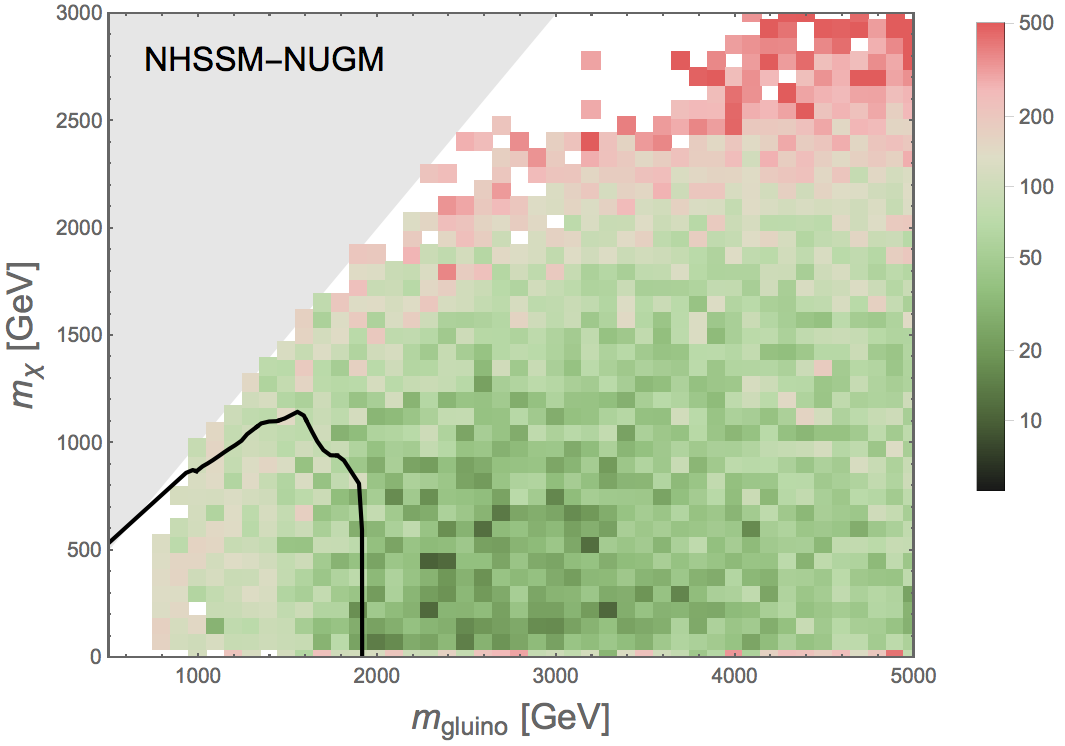}
\includegraphics[width=0.49\linewidth]{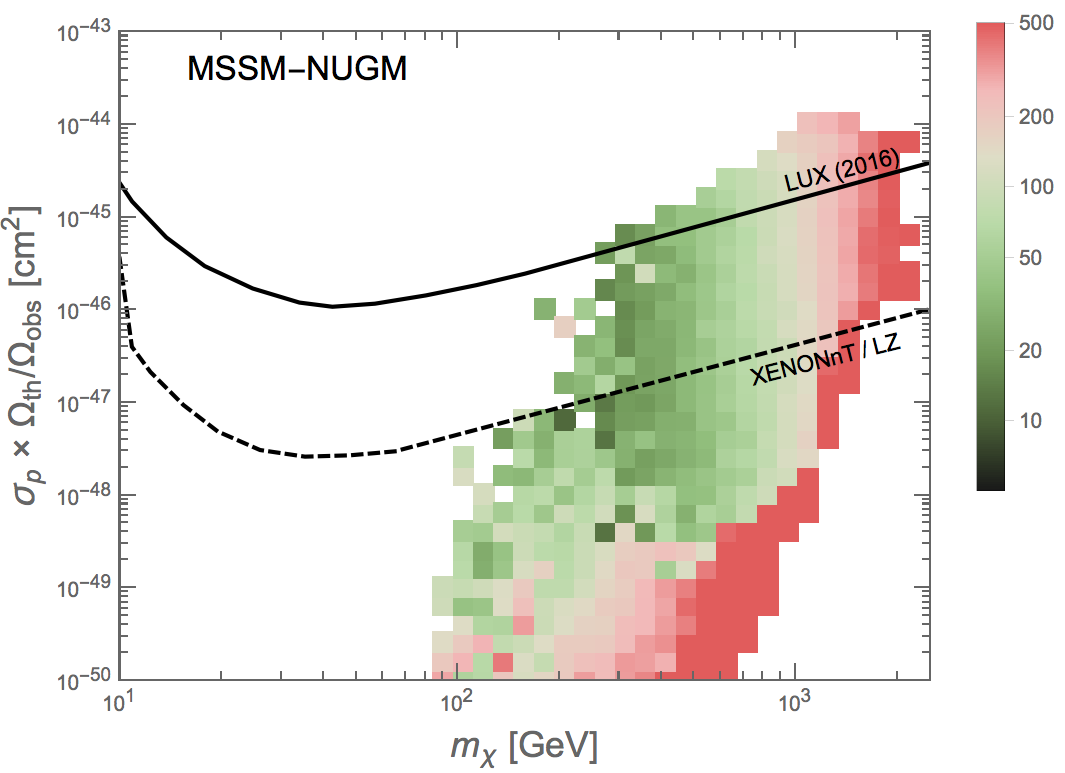}\hfill 
\includegraphics[width=0.49\linewidth]{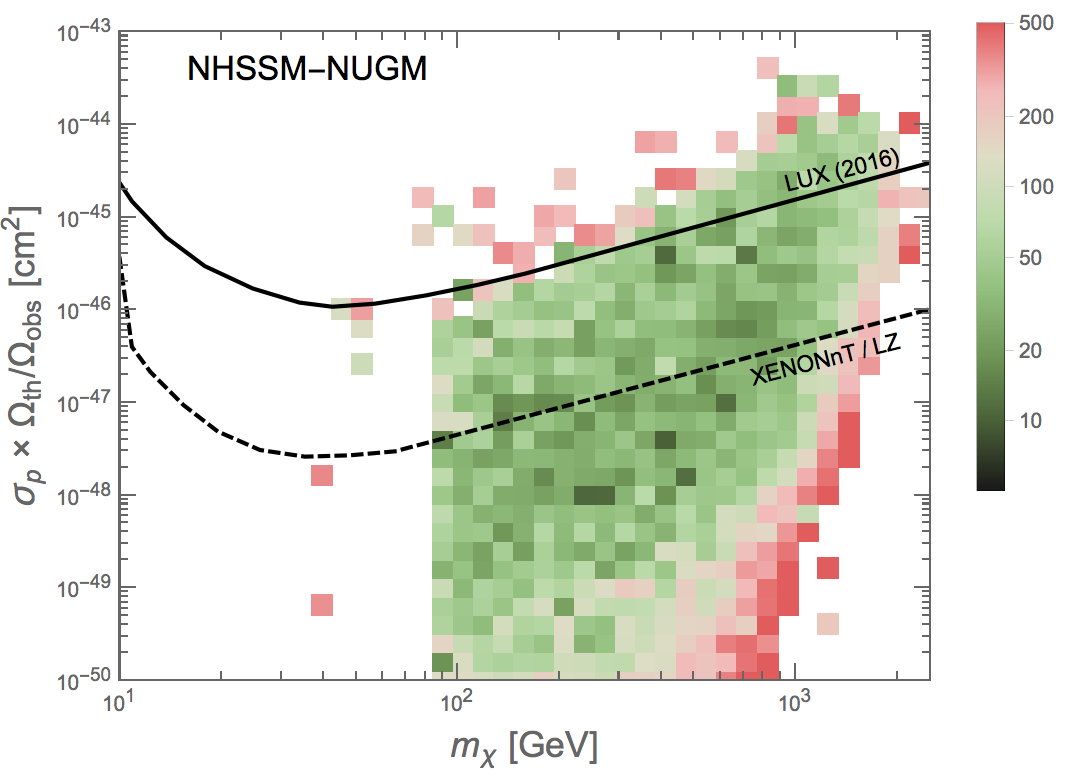}
\caption{Same as Fig.~\ref{fig:CMSSM_gluino} but for the case of non-universal gaugino masses.} 
\label{fig:MSSMNUGM_gluino}
\end{figure}
Let us finally discuss how future collider and direct detection experiments test the natural range of parameters for the case of non-universal gaugino masses.
In Fig.~\ref{fig:MSSMNUGM_gluino} we again show the FT in the in the $m_\chi  - m_{\tilde{g}}$ plane to estimate the sensitivity of future collider searches (top panel)
as well as the FT in the $m_\chi  - \sigma_p$ plane (bottom panel). Interestingly, once we impose the mass of the SM-like Higgs boson to be in the measured range, the
smallest fine tuning is found for gluino masses above the current bounds and the region with small FT extends to very large gluino masses.
In fact for gluino masses as heavy as 5 TeV the fine tuning can still be between 20 and 30, as can also be inferred from Tab~\ref{tab:FT}.
This observation applies both to the case with and without the extra non-holomorphic contribution to the Higgsino mass, although the case with extra terms 
is slightly preferred. What is different however is the dependence on the LSP mass, as expected: 
Without the extra contribution there is a large FT penalty when going to large LSP masses, while with this contribution the increase in fine tuning is very mild. 
Turning to the dark matter properties of these models, it can be seen that the region with small FT extends down to very small direct detection cross section. Indeed the cross sections are often below the expected sensitivity of LZ and XENONnT and can be as small as the neutrino background so that it seems impossible to probe the entire natural parameter space of the model.

To summarise we present in Tab.~\ref{tab:FT} the minimum FT $\Delta_\text{min}$ for all considered cases after applying the various cuts discussed in the text,
both for the current bounds and estimating the sensitivity of future experiments. 
\begin{table}[hbt]
\begin{center}
\begin{tabular}{|c|ccc|cccc|}
\hline 
\multicolumn{8}{|c|}{CMSSM}\\
\hline
&\multicolumn{3}{|c|}{current} & \multicolumn{4}{|c|}{prospects}\\
\hline
Cut            & LHC \& Higgs & soft DM & strong DM & $m_{\tilde{g}} \geq 3$~TeV & DD & $m_{\tilde{g}} \geq 5$~TeV & DD\\  
\hline
$\Delta_\text{min}$  & 134 & 216  & 276     & 231 & 271 & 686 & - \\
\hline 
\hline 
\multicolumn{8}{|c|}{CNHSSM}\\
\hline
&\multicolumn{3}{|c|}{current} & \multicolumn{4}{|c|}{prospects}\\
\hline
Cut            & LHC \& Higgs & soft DM & strong DM & $m_{\tilde{g}} \geq 3$TeV & DD & $m_{\tilde{g}} \geq 5$~TeV & DD\\ 
\hline
$\Delta_\text{min}$ & 114 & 116  & 166      & 227 & 264 & 665 & 677 \\
\hline
\hline
\multicolumn{8}{|c|}{MSSM-NUHM}\\
\hline
&\multicolumn{3}{|c|}{current} & \multicolumn{4}{|c|}{prospects}\\
\hline
Cut            & LHC \& Higgs & soft DM & strong DM & $m_{\tilde{g}} \geq 3$~TeV & DD & $m_{\tilde{g}} \geq 5$~TeV & DD\\ 
\hline
$\Delta_\text{min}$ & 160 & 302  & 501     & 292  & 617 & 702 & 1406 \\
\hline 
\hline
\multicolumn{8}{|c|}{NHSSM-NUHM}\\
\hline
&\multicolumn{3}{|c|}{current} & \multicolumn{4}{|c|}{prospects}\\
\hline
Cut            & LHC \& Higgs & soft DM & strong DM & $m_{\tilde{g}} \geq 3$~TeV & DD & $m_{\tilde{g}} \geq 5$~TeV & DD\\ 
$\Delta_\text{min}$ & 161 & 307 & 368     & 272 & 626 &  698 & 1381 \\
\hline 
\hline 
\multicolumn{8}{|c|}{MSSM-NUGM}\\
\hline
&\multicolumn{3}{|c|}{current} & \multicolumn{4}{|c|}{prospects}\\
\hline
Cut            & LHC \& Higgs & soft DM & strong DM & $m_{\tilde{g}} \geq 3$~TeV & DD & $m_{\tilde{g}} \geq 5$~TeV & DD\\ 
\hline
$\Delta_\text{min}$  & 11 & 11  & 117      & 17 & 17 & 29 & 29\\
\hline 
\hline 
\multicolumn{8}{|c|}{NHSSM-NUGM}\\
\hline
&\multicolumn{3}{|c|}{current} & \multicolumn{4}{|c|}{prospects}\\
\hline
Cut            & LHC \& Higgs & soft DM & strong DM & $m_{\tilde{g}} \geq 3$~TeV & DD & $m_{\tilde{g}} \geq 5$~TeV & DD\\ 
\hline
$\Delta_\text{min}$  & 10 & 10 & 23       & 11 & 11 & 23 & 23 \\
\hline
\end{tabular}
\end{center}
\caption{The best fine-tuning $\Delta_\text{min}$ found after applying the different cuts in the MSSM for all considered boundary conditions. The first three columns correspond to the current limits with cuts as described in the text. The last four columns are sensitivity estimates for a lower bound on the gluino mass $m_{\tilde{g}}$ of 3 and 5 TeV respectively. The columns denoted by DD in addition take into account a bound on the LSP nucleon cross section as expected for the future direct detection experiments XENONnT and LZ.}
\label{tab:FT}
\end{table}

\section{Summary and Conclusions}
\label{sec:conclusions}
In this work we have reassessed the FT in the MSSM for a number of different GUT scale SUSY breaking boundary conditions. Specifically we considered the fully constrained MSSM with and without an additional non-holomorphic Higgsino mass term. We further relaxed the condition of universality of Higgs and gaugino masses at the GUT scale, see Tab.~\ref{tab:BC} for all considered boundary conditions. We find that due to the proper inclusion of loop corrections the FT is typically smaller by about a factor of two
compared to a tree-level analysis of the sensitivity measure for the well-known case of the CMSSM. The inclusion of the non-holomorphic Higgsino mass does not appreciably change the FT for this case, but allows for more flexibility in the neutralino sector, making it possible to satisfy the dark matter relic abundance constraint much more easily. Non-universal Higgs masses don't improve on the FT - in fact the FT is typically even worse due to the loss of the scalar focus point.
The situation is very different for non-universal gaugino masses, which allow for a significant reduction in FT, assuming that the ratio of gaugino masses is fixed
within the underlying theory. In this case we find points with FT as small as 10 which pass all the collider and dark matter limits, if one allows for a dark matter
underabundance. If we require the correct dark matter density, which typically requires a Higgsino LSP with mass of about 1 TeV, the FT increases significantly, 
becoming larger than 100 due to the large $\mu$ term when no additional contribution to the Higgsino mass is taken into account. With such an additional contribution
however the FT can stay as small as $\sim 20$ even for a 1 TeV Higgsino.

We also evaluated the prospects of testing the naturalness of these different models within the next 10-20 years, taking into account more stringent
cuts on the gluino mass as well as future direct detection experiments. We find that while the fully universal case as well as the case with non-universal Higgs
masses will be significantly more constrained, the case of non-universal gauginos allows for very large gluino masses without a large penalty on the FT measure.
In fact when taking into account the known SM-like Higgs mass, the region with smallest FT is for gluino masses above current bounds.
Even taking into account future direct detection experiments such as XENONnT and LZ, these low fine tuned regions survive, offering an interesting way to evade all
bounds while still maintaining one of the original motivations of SUSY, i.e. fully solving the hierarchy problem of the SM.

\section*{Acknowledgements}
This work is supported by the German Science Foundation (DFG) under the Collaborative 
Research Center (SFB) 676 Particles, Strings and the Early Universe as well as the ERC Starting Grant `NewAve' (638528).

\bibliography{NMSSM}
\bibliographystyle{ArXiv}

\end{document}